\documentclass[twocolumn,times]{aastex631}

\usepackage[english]{babel}
\usepackage[utf8]{inputenc}
\usepackage{graphicx}
\usepackage{gensymb}
\usepackage{CJKutf8}
\usepackage{comment}
\usepackage[T1]{fontenc}
\usepackage{upgreek}

\received{}
\revised{}
\accepted{\today}
\submitjournal{APJ}

\newcommand{\snapshotAAmulti}{ULX4a-MG} % Snapshot 08250 multi-group
\newcommand{\snapshotBmulti}{ULX2.5-MG} % Snapshot 18056 multi-group
\newcommand{\snapshotAA}{ULX4a} % Snapshot 08201
\newcommand{\snapshotAB}{ULX4b} % Snapshot 08481
\newcommand{\snapshotB}{ULX2.5} % Snapshot 18000
\newcommand{\snapshotC}{ULX1.3} % Snapshot 15900
\newcommand{\nsamp}{N_{\rm s}}
\newcommand{\nphot}{N_{\rm ph}}
\newcommand{\ncell}{N_{\rm cell}}
\newcommand{\rg}{$r_{\rm g}$}
\newcommand{\Athena}{\texttt{Athena++}}
\begin{document}
\title{Spectral calculations of 3D RMHD simulations of super-Eddington accretion onto a stellar-mass black hole}

\author[0000-0003-0148-2817]{Brianna S. Mills}
\email{bri@virginia.edu}

\author[0000-0001-7488-4468]{Shane W. Davis}
\affiliation{Department of Astronomy, University of Virginia, 530 McCormick Road Charlottesville, VA 22904, USA}

\author[0000-0002-2624-3399]{Yan-Fei Jiang (\begin{CJK*}{UTF8}{gbsn}姜燕飞\end{CJK*})}
\affiliation{Center for Computational Astrophysics, Flatiron Institute, New York, NY 10010, USA}

\author[0000-0002-8183-2970]{Matthew J. Middleton} 
\affiliation{Department of Physics and Astronomy, University of Southampton, Highfield, Southampton SO17 1BJ, UK}

%%%%%%%%%%%%%%%%%%%%%%%%%%%%%%%%%%%%%%%%%%%%%%%%%%%%%%%%%%%%%%%%%%%%%%%%%%%%%%%%%%%%%%%%%%%%%
%%%%%%%%%%%%%%%%%%%%%%%%%%%%%%%%%%     ABSTRACT    %%%%%%%%%%%%%%%%%%%%%%%%%%%%%%%%%%%%%%%%%% %%%%%%%%%%%%%%%%%%%%%%%%%%%%%%%%%%%%%%%%%%%%%%%%%%%%%%%%%%%%%%%%%%%%%%%%%%%%%%%%%%%%%%%%%%%%%
\begin{abstract}
We use the Athena++ Monte Carlo (MC) radiation transfer module to post-process simulation snapshots from non-relativistic Athena++ radiation magnetohydrodynamic (RMHD) simulations. These simulations were run using a gray (frequency integrated) approach but were also restarted and ran with a multi-group approach that accounts for Compton scattering with a Kompaneets operator. These simulations produced moderately super-Eddington accretion rates onto a 6.62 $M_\odot$ black hole. Since we only achieve inflow equilibrium out to 20-25 gravitational radii, we focus on the hard X-ray emission. We provide a comparison between the MC and RMHD simulations showing that the treatment of Compton scattering in the gray RMHD simulations underestimates the gas temperature in the regions above and below the accretion disk.  In contrast, the restarted multi-group snapshots provides a treatment for the radiation field that is more consistent with the MC calculations, and result in post-processed spectra with harder X-ray emission compared to their gray snapshot counterparts.  We characterize these MC post-processed spectra using commonly employed phenomenological models used for spectral fitting. We also attempt to fit our MC spectra directly to observations of the ultraluminous X-ray source (ULX) NGC 1313 X-1, finding best fit values that are competitive to phenomenological model fits, indicating that first principle models of super-Eddington accretion may adequately explain the observed hard X-ray spectra in some ULX sources.
\end{abstract}

%%%%%%%%%%%%%%%%%%%%%%%%%%%%%%%%%%%%%%%%%%%%%%%%%%%%%%%%%%%%%%%%%%%%%%%%%%%%%%%%%%%%%%%%%%%%%
%%%%%%%%%%%%%%%%%%%%%%%%%%%%%%%%%     INTRODUCTION    %%%%%%%%%%%%%%%%%%%%%%%%%%%%%%%%%%%%%%% %%%%%%%%%%%%%%%%%%%%%%%%%%%%%%%%%%%%%%%%%%%%%%%%%%%%%%%%%%%%%%%%%%%%%%%%%%%%%%%%%%%%%%%%%%%%%
\section{Introduction}
\label{s-introduction}

Ultra-luminous X-ray sources (ULXs) are point-like, off-nuclear extragalactic objects observed to have X-ray luminosities comparable to or in excess of the critical Eddington luminosity $L_{\rm X }\gtrsim10^{39}$ erg/s (assuming isotropic emission for a $10 M_\odot$ black hole; see \citealp{pinto_walton2023, KLM2023} for a review of ULXs).  The majority of ULXs are now accepted to be X-ray binary systems with super-Eddington rates of accretion onto a compact object, namely a stellar mass $M<100 M_\odot$ black hole \citep{Poutanen2007, Middleton2015} or neutron star \citep{skinneretal1982,bachettietal2014}. Some small fraction of ULXs may yet harbor sub-Eddington accretion rates onto intermediate-mass black holes $M\gtrsim100 M_\odot$ (IMBHs; \citealp{farrelletal2009, mezcuaetal2013, earnshaw2016, brightmanetal2016, webbetal2017, oskinovaetal2019}). The physical mechanisms which drive super-Eddington accretion are still under investigation and require numerical simulations in order to evaluate existing models of black hole accretion.

The classical picture of an optically thick, geometrically thin accretion disk \citep{shakura_sunyaev1973} is used to model black hole X-ray binaries (BHXBs) and is a generally applicable when the accretion rate is sub-Eddington ($L/L_{\rm{Edd}}<0.3$) where the disk geometry (defined by the disc semi-thickness, H and radius, R) remains thin $H<<R$.  If ULXs are indeed IMBHs, the spectra are expected to resemble scaled up versions of BHXB spectra, showing cooler accretion disks as the black hole mass increases (e.g. \citealp{milleretal2004}).  

Observations of ULXs typically show a soft, thermal X-ray component and a hard thermal component with a rollover below $\sim10$ keV \citep{gladstoneetal2009,bachettietal2014}, the latter supporting the interpretation of super-Eddington accretion. Early models debated whether this hard X-ray emission originated from coronal emission from IMBHs \citep{milleretal2004} or Comptonized emission from super-Eddington accretion \citep{gladstoneetal2009,socrates_davis2006}. However, classically, one would expect the innermost regions to have a different spectral shape due to optical depth effects and anisotropy \citep{Poutanen2007}.

Super-Eddington accretion is expected to deviate from the classical \cite{shakura_sunyaev1973} thin disk approximation, as the radiation pressure exceeds gravity.  Processes like advection \citep{abramowiczetal1988} and radiatively driven outflows \citep{shakura_sunyaev1973,ohsuga_mineshige2011} may reduce the radiative efficiency and result in geometrically thicker flows in the super-Eddington regime. Advection can directly affect the observed spectra \citep{straubetal2011,kubota2019}. Strong optically thick winds are also expected to be launched in these systems (and widely detected in ULXs: \citealp{Middleton2014, Middleton2015b, Pinto2016, Pinto2020, Walton2016, Kosec2021}), which likely shroud the outer accretion disk and can contribute additional low energy flux for preferential sight lines. 

Due to the complex nature of describing three-dimensional super-Eddington accretion flows, numerical simulations are a key tool for studying this regime.  Several radiation hydrodynamic (RHD; \citealp{ohsugaetal2005}), radiation magentohydrodynamic (RMHD; \citealp{ohsuga_mineshige2011,jiangetal2014}), and general relativistic RMHD (GRRMHD; \citealp{mckinneyetal2014,fragileetal2014,sadowskietal2015,sadowski_narayan2016}) simulations have been performed to understand the physical mechanisms involved in super-Eddington accretion. In these simulations, the radiation transfer equation is often integrated over frequency (the ``gray'' approximation) to reduce the computational expense. In many case, the angle-integrated radiation moments (e.g. radiation flux and/or energy density) are solved for, which usually requires a closure relation (e.g. flux limited diffusion, \citealp{turnerstone2001, howellgreenough2003, krumholzetal2007,moensetal2022}; M1 closure, \citealp{levermore1984,gonzalezetal2007,skinneretal2013,wibkingkrumholz2022}; or variable Eddington tensor method, \citealp{jiangetal2012,davisetal2012,jiangetal2014,asahinaetal2020,menonetal2022}) to complete the the radiation moments.  An alternative approach, which is used for the simulations discussed in this work, is the direct solution of the gray radiation transfer equation \citep{stoneetal1992,jiangetal2014,jiang2021}, which is then coupled to the fluid by computing the radiative cooling/heating and radiation force.

There have been significant efforts to simulate global accretion flows in the vicinity of black holes and utilize them to generate synthetic observables to compare with observations.  Perhaps the most impactful is the effort by the Event Horizon Telescope to interpret the very long baseline interferometric images of M87* and Sgr A* \citep{eht2019,eht2022}.  In these systems, the flows are relatively optically thin to electron scattering and the modeling of Compton scattering is not essential to primary the imaging effort.  Our current study is focused on more radiatively efficient and optically thick flows where electron scattering opacity dominates.  Previous work includes efforts to generate spectra from GRMHD simulations that utilized simple cooling prescriptions to keep the disk thin\citep{zhuetal2012,schnittmanetal2013,kinchetal2019,kinchetal2021} to study the sub-Eddington or near Eddington regime, non-relativistic radiation hydrodynamics simulations of super-Eddington accretion \citep{kawashima2012,kitaki2017}, and radiative GRRMHD simulations of the super-Eddington regime \citep{narayanetal2017}.  

Spectral post-processing is commonly performed using Monte Carlo radiation transfer methods, which are useful for modeling the effects of Compton scattering. MC methods such as \texttt{GRMONTY} \citep{dolence2009}, \texttt{Pandurata} \citep{schnittman2013}, or \texttt{RAIKOU} \citep{2021arXiv210805131K} model Compton scattering and include general relativistic effects. The \texttt{HEROIC} code \citep{narayanetal2017} provides similar capabilities, but uses a combination of short and long characteristics instead of MC.  These can also be coupled to photoionization calculations to produce predictions for atomic features, such as the Fe K$\alpha$ line \citep{kinchetal2019}. The inclusion of Compton scattering is a key ingredient because it dominates the thermodynamic coupling between the radiation and gas near or above the photosphere \citep{narayanetal2017,kinchetal2020}.

In this work, we use the MC radiation transfer module in \Athena{} to post-process \Athena{} RMHD simulation snapshots and aim to describe these results with current black hole accretion models, as well as compare the simulated spectra to data for the ULX NGC 1313 X-1. Although the simulations performed here rely primarily on the non-relativistic gray RMHD module, two recent developments to \Athena{} offer potential improvements for future work.  The first is a multi-group implementation\citep{jiang2022} that facilitates multi-frequency transfer and better treatment of Compton heating and cooling.  The second is a fully general relativistic formalism \citep{white2023}.  As we discuss in Section \ref{s-results:multigroup}, we utilize the multi-group method in this work to obtain a more accurate estimate for the temperature distribution in the current simulations. Spectral calculations with the GR implementation will be a focus of future work. 

The plan of this work is as follows:  In Section \ref{s-methods} we discuss the MC and \Athena{} methods used in our spectral post-processing analysis.  In Section \ref{s-results} we present the gray and multi-group RMHD spectral analysis results, along with image results and a comparison to phenomenological spectral models and fits to the spectrum of NGC 1313 X-1.  We discuss the caveats, implications, and comparison of our results to previous work in Section \ref{s-discussion}. Finally, we summarize the key points of this work in Section \ref{s-summary}.

%%%%%%%%%%%%%%%%%%%%%%%%%%%%%%%%%%%%%%%%%%%%%%%%%%%%%%%%%%%%%%%%%%%%%%%%%%%%%%%%%%%%%%%%%%%%%
%%%%%%%%%%%%%%%%%%%%%%%%%%%%%%%%     METHODS    %%%%%%%%%%%%%%%%%%%%%%%%%%%%%%%%%%%%%%%%%%%%% %%%%%%%%%%%%%%%%%%%%%%%%%%%%%%%%%%%%%%%%%%%%%%%%%%%%%%%%%%%%%%%%%%%%%%%%%%%%%%%%%%%%%%%%%%%%%
\section{Methods}
\label{s-methods}
We utilize the \Athena{} code \citep{jiang2019super,stoneetal2020,whiteetal2016} in two configurations -- using the \Athena{} RMHD simulation snapshots of super-Eddington accretion onto a 6.62 $M_\odot$ black hole, and using the \Athena{} Monte Carlo radiative transfer module (Davis et al. in prep) to post-process the snapshots.  Here we describe both configurations separately, and discuss the methods used for post-processing in the last subsection.
%**********************************************************************************
%*********************    MC Radiation Transfer Code    ***************************
%**********************************************************************************
\subsection{MC radiation transfer code}
\label{s-methods:montecarlo}

The standard \Athena{} RMHD simulations utilize gray opacities (frequency averaged opacities) and thus do not directly provide any spectral information. To extract frequency information needed to produce the spectra, we utilize the \Athena{} Monte Carlo (MC) radiation transfer module (\citealp{davisetal2009}, Davis et al. in prep) to compute the radiation field throughout an \Athena{} simulation snapshot.  The MC module  utilizes the \Athena{} code structure and mesh, allowing it to be run concurrently with the simulations.  It can also be utilized to read in output simulation snapshot for post-processing, which is how it is used here.  Although the module can be used to perform MC transfer on the full three-dimensional refined simulation mesh, we focused here on two dimensional axisymmetric calculations, where finer/coarser levels are prolongated/restricted to an intermediate refinement level uniform mesh.  The MC calculation proceeds by creating and then tracking photon samples throughout the mesh. The samples (often referred to as photon packets or superphotons) can be viewed as statistical ensembles of a large number of photons with common properties. These properties of the photons are initialized and evolved using pseudorandom numbers to draw from distributions in positions, photon energies, scattering angles, etc. until they are either absorbed or leave the domain.

In this work we model free-free emission and absorption and unpolarized Compton scattering as the primary radiative processes. Each photon sample has a statistical weight corresponding to the number of photons in the packet.  We model emission by randomly sampling each zone and assigning a weight corresponding to the volume integrated free-free emissivity from the sampled cell.  We assume a total number of photon samples $\nsamp$ and $\ncell$ cells in the mesh.  If we label cells by index $i$ and photons samples with index $j$, the total number of physical photons emitted in cell $i$ can be written
\begin{equation}
N_i = \int \frac{j(\nu,T_i,\rho_i)}{h\nu} d \Omega d\nu \mathcal{V}_i  \Delta t_{\rm int},
\end{equation}
Here, $\Omega$ is the solid angle, $\mathcal{V}_i$ is the volume of cell $i$, $j(\nu,T_i,\rho_i)$ is the free-free emissivity as function of temperature and density within the cell \citep{1979rpa..book.....R}, and $\Delta t_{\rm int}$ is the (arbitrary) integration time interval. The statistical weights are defined so that
\begin{equation}
\sum_{j=1}^{\nsamp} w_j = \sum_{i=1}^{\ncell} N_i = \nphot,
\label{eq:nphotons1}
\end{equation}
where $\nphot$ is the total number of physical photons emitted within the entire mesh. We can define the probability $P_i$ for a photon to be emitted in zone $i$ as $P_i = N_i/\nphot = 1/\ncell$.  Then, the average number of photon samples emitted in cell $i$ is $P_i \nsamp$, and we have
\begin{equation}
w_i = \frac{N_i \ncell}{\nsamp}.
\end{equation}
This procedure yields photon weights that can differ by orders of magnitude.  This is often frowned upon in the MC literature because more uniform weighting is generally variance reducing.  We have, however, also implemented an equal weighting scheme where the initial cells of photon samples are chosen proportional to their volume weighted emissivity and found this scheme ultimately results in larger statistical errors in our output spectra per computational second when compared with the scheme used here. This is primarily due to the large scattering optical depths to escape for photons launched in the highest emission cells (Davis et al. in preparation).

Finally, the direction of the photon is randomly sampled from an isotropic distribution, and the energy of the photon is drawn from a log normal distribution in photon energy.  We then further adjust the weight so that binned photons match free-free distribution in photon frequency. Photon movement is handled in the Eulerian (coordinate) frame, while emission, scattering, and absorption occur in the comoving fluid frame. Photon sample properties are Lorentz boosted between the coordinate and fluid (comoving) frame for these interactions.
 
Photon samples are moved between scattering/absorption events by drawing an exponentially distributed dimensionless path length $\tau$ to the next absorption/scattering event via $\tau= -\ln \xi$, where $\xi$ is a pseudorandom number uniformly distributed in the interval (0,1).  This dimensionless path length can be thought of as the optical depth to the next scattering/absorption event, and is computed as a series of steps $l_k$ (enumerated with subscript $k$) so that 
 \begin{equation}
\tau = \sum_{k} \frac{l_k}{\alpha_{\nu,k} +\sigma_{\nu,k}},
 \end{equation}
where $\alpha_\nu$ is the absorption extinction coefficient, and $\sigma_\nu$ is the scattering extinction coefficient.  The scattering and absorption coefficients are the products of the corresponding opacities and density, which are evaluated in the comoving frame and then boosted to the Eulerian frame. In the scheme used here each step $k$ represents a movement of the photon sample to the location of the next scattering/absorption event or the nearest cell face, whichever comes first. This continues until the requisite value of $\tau$ is reached or the photon sample escapes the domain.  Photon samples are assumed to travel along straight lines, but we use a spherical mesh, so that computing where the photon sample leaves the current cell requires solving quadratic relations and accounting for possible turning points in $r$ and $\theta$ (Davis et al., in preparation).
 
Each interaction of photon sample with matter results in a combination of absorption and scattering, which is handled by reductions in $w$. We have $w'=w \epsilon$, where $w'$ is the new weight after scattering and 
\begin{equation}\label{eq:absorption}
    \epsilon = \frac{\alpha_\nu}{\alpha_\nu + \sigma_\nu}.
\end{equation}
If the statistical weight falls below a small threshold value (based on the initial emissivity), the photon is considered absorbed and further evolution is terminated. The outgoing photon energy and direction after Compton scattering follow from procedures described in \citet{1983ASPRv...2..189P}, except that we tabulate the scattering cross section using a method similar to that described in \citet{dolence2009}.

When photons escape through the domain boundary, their energies, locations, and angles are tabulated in a photon list output that is then used to generate spectra.  The MC calculation also tabulates cell-averaged radiation moments such as the energy density, radiation flux vector, and pressure tensor, as well as user defined quantities such as the net radiative cooling, average photon energy, and average energy mean opacity in each cell.  These are output in standard \Athena{} formats, such as HDF5 and VTK.

%**********************************************************************************
%*********************    Athena++ RMHD SIMULATIONS    ***************************
%**********************************************************************************

%------------------------------  Table  ----------------------------------
\begin{deluxetable}{ccccc}[ht!]
\tablecaption{\Athena{} RMHD simulation snapshots}
\tablecolumns{5}
\tablehead{\colhead{Snapshot} & \colhead{$\langle \dot{M} \rangle/\dot{M}_{\rm Edd}$} & \colhead{$\theta_{\rm f}$} & \colhead{$L_{\rm f}$ (erg $\rm{cm}^{-2} \rm{s}^{-1}$}) & \colhead{$\eta_{\rm f}$} }
\startdata
\snapshotAA{}   & -4.15  & $37^{\circ}$  &  1.03e+39  & 2.56\% \\
\snapshotAB{}   & -3.93  & $37^{\circ}$  &  8.77e+38  & 2.29\% \\
\snapshotB{}    & -2.53  & $50^{\circ}$  &  2.78e+38  & 1.13\% \\
\snapshotC{}    & -1.31  & $55^{\circ}$  &  1.51e+38  & 1.18\% \\
\hline
\snapshotAAmulti{} &  -4.02  &  $37^{\circ}$  & 1.31e+39  & 3.34\% \\
\snapshotBmulti{}  &  -2.53  &  $50^{\circ}$  & 4.73e+38  & 1.92\% \\
\enddata
\tablecomments{\Athena{} RMHD simulation snapshots of a 6.62$M_{\odot}$ black hole used in this analysis. All snapshots are azimuthally averaged and are limited to the inner 25\rg{}. The first column corresponds to the ratio of the radially averaged mass accretion rate $\langle \dot{M} \rangle$ in terms of the Eddington mass accretion rate $\dot{M_{\rm Edd}}$.  The negative sign indicates accretion towards the black hole.  The second column corresponds to the polar funnel angle $\theta_{\rm f}$, the opening angle relative to the polar axis representing the approximate boundary between the funnel region and the accretion disk. Photons emerging from this polar funnel angle are collected for spectral post-processing and have corresponding funnel luminosities $L_{\rm f}$.  The last column is the calculated radiative efficiency $\eta_{\rm f}$ of the funnel region.  Snapshots \snapshotAA{} and \snapshotAB{} were taken from the same simulation run (at different times), whereas Snapshots \snapshotB{} and \snapshotC{} are independent simulation runs.  The snapshots with the suffix ``-MG'' correspond to the two gray simulations chosen for the multi-group RMHD implementation \citep{jiang2022}.}
\end{deluxetable}\label{tab:parameters}
%--------------------------------------------------------------------------//

\subsection{Athena++ RMHD simulation snapshots}
\label{s-methods:simulations}

\Athena{} has been rewritten in C++ compared to its predecessor, Athena \citep{stoneetal2008}.  \Athena{} now includes adaptive mesh refinement \citep{stoneetal2020} and special and general relativistic capabilities \citep{whiteetal2016,white2023}.  In the current work, however, a pseudo-Newtonian potential is used to mimic the effects of general relativity around a Schwarzschild black hole \citep{paczynskywiita1980}.  Results from a GRRMHD implementation of \Athena{} and subsequent spectra will be reported in future work.

We performed a series of global, three-dimensional RMHD simulations for a 6.62 $M_{\odot}$ black hole accreting at several super-Eddington mass accretion rates assuming a 10\% radiative efficiency so that $\dot{M}_{\rm{Edd}} \equiv 10 L_{\rm{Edd}}/c^2$.  We used the explicit integration RMHD module in \Athena{}, which uses an algorithm similar to \citet{jiangetal2014}, but with updates that solve a radiation transfer equation of the form presented in \citet{jiang2021}. The simulation setup for these snapshots is similar to the setup described in \citet{huangetal2023}, where the ideal MHD equations are coupled with the time-dependent radiation transfer equation (see \citealp{jiangetal2014} equations 1-4, and \citealt{jiang2021} equations 4-6).  A rotating gas torus was initialized in hydrostatic equilibrium and threaded with toroidal magnetic fields.  Accretion onto the black hole happens via the magnetorotational instability \citep{balbushawley1991} and the mass accretion rate is varied for each simulation based on the initial magnetic field configuration \citep[see e.g.,][]{huangetal2023}.

%------------------------------  Figure  ----------------------------------
\begin{figure}[t!]
    \centering
    \includegraphics[width=\columnwidth]{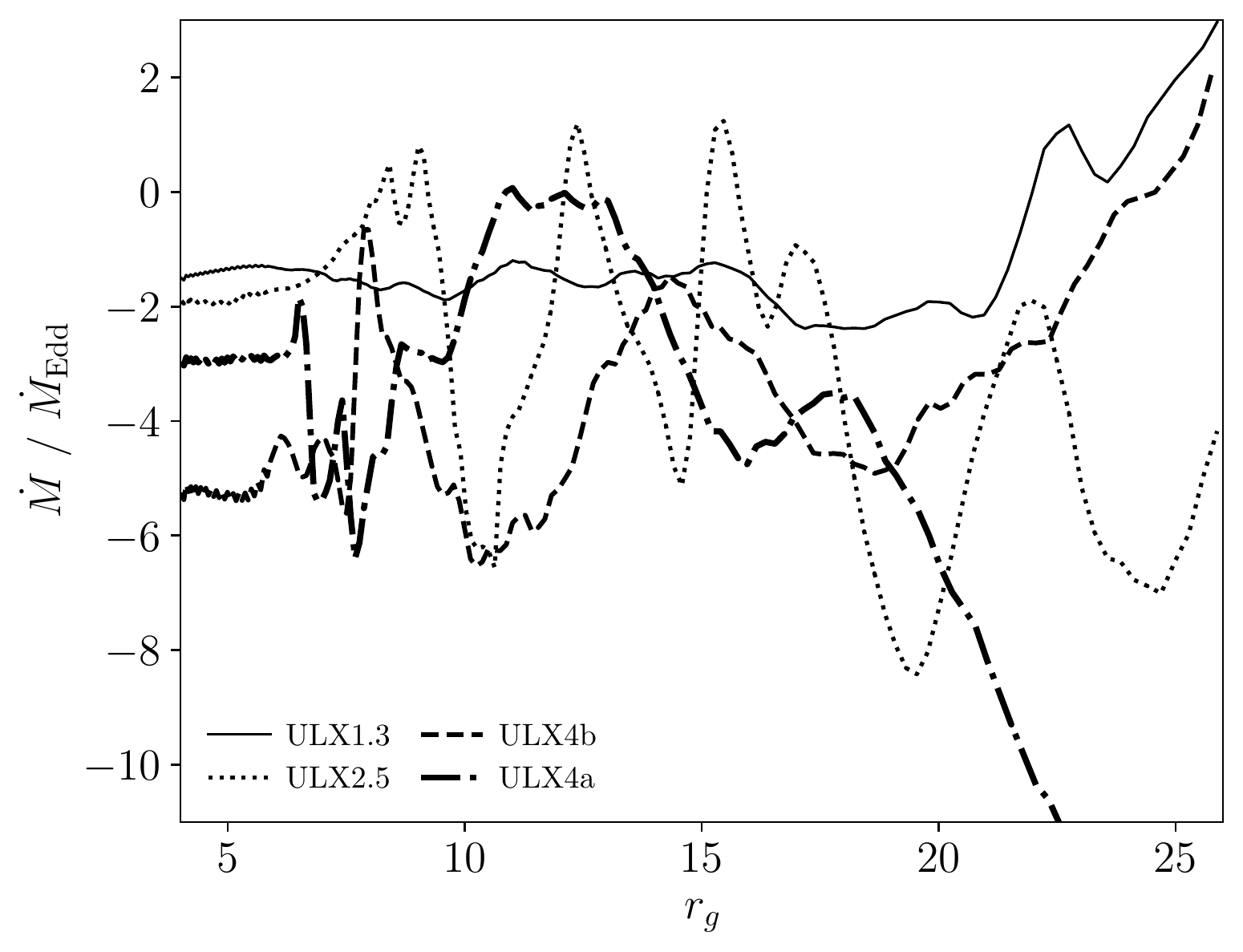}
    \caption{The mass accretion rate $\dot{M}$ as a function of gravitational radius \rg{} from the black hole for each \Athena{} RMHD simulation snapshot (see Table \ref{tab:parameters}).  \snapshotAA{} is the dot-dash line, \snapshotAB{} is the dashed line, \snapshotB{} is the dotted line, and \snapshotC{} is the solid line.}
    \label{fig:mdot}
\end{figure}
%----------------------------------------------------------------------------//

The simulations self-consistently form an accretion disk and reach a quasi-steady state for the inner disk.  Figure \ref{fig:mdot} shows the mass accretion rate in terms of $\dot{M}_{\rm Edd}$ for a 6.62$M_\odot$ black hole as a function of radius within the inner 25\rg{} where $r_{\rm{g}} = GM/c^2$ is the gravitational radius. The mass accretion rates are relatively steady-state within 25\rg{} of the black hole, but not at larger radii. The four snapshots and their radially-averaged mass accretion rates (over the $25$\rg{}) are listed in Table \ref{tab:parameters}.  Snapshots \snapshotAA{} and \snapshotAB{} have nearly the same mass accretion rate ($\dot{M} \simeq -4 \dot{M}_{\rm Edd}$) and are both from the same simulation run (at different times), thus we named them \snapshotAA{} and \snapshotAB{}.  Snapshot \snapshotB{} and Snapshot \snapshotC{} are independent simulation runs with average mass accretion rates $\dot{M} {\simeq} -2.5$ and $-1.3 \dot{M}_{\rm Edd }$, respectively.

Although the \Athena{} RMHD calculations have adaptive mesh refinement capabilities, the MC code works most efficiently on a uniform grid. For efficient parallelization, we chose one uniform refinement level for our analysis. We selected an appropriate refinement level such that all snapshot grids were approximately the same size 256 x 128 x 256 cells in $r$, $\theta$, and $\phi$ (respectively). The accretion disk located in the inner 25\rg{} roughly corresponds to the 80 innermost zones in radius at this level, and covers a range of $\theta$ from 0 to $\pi$, and a range of $\phi$ from 0 to $2\pi$. Due to the approximately axisymmetric nature of the simulations, we chose to azimuthally average each snapshot for our post-processing analysis.  This has little effect on the output spectra, but greatly improves the statistics for cell averaged quantities, examples of which are presented in Figures 3-6.

%------------------------------  Figure  ----------------------------------
\begin{figure}[t!]
    \centering
    \includegraphics[width=\columnwidth]{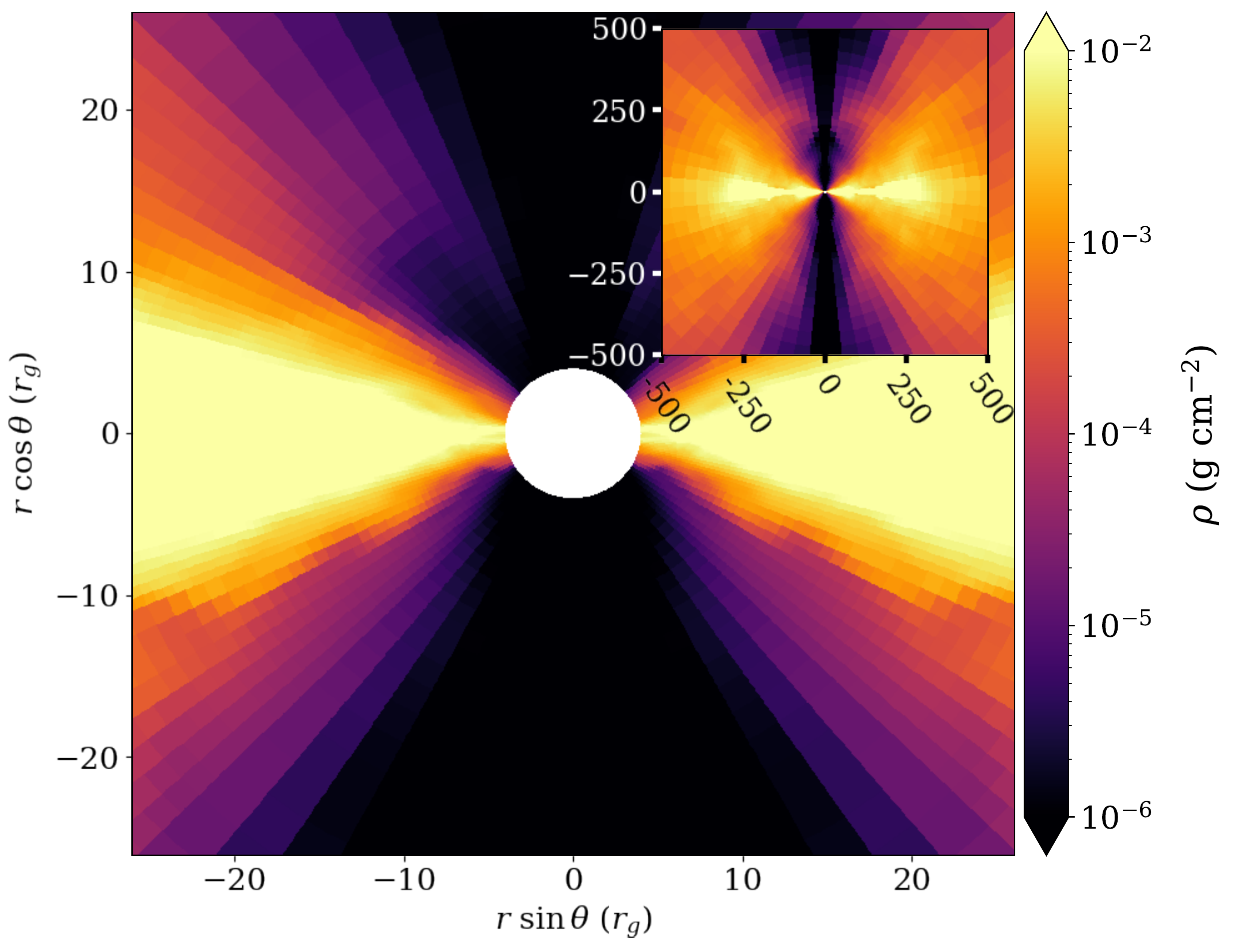}
    \caption{Azimuthally averaged gas density $\rho$ (g cm$^{-2}$) of the \Athena{} simulation showing the inner 25 \rg{} with a small inset plot that shows the gas density out to 500 \rg{}.}
    \label{fig:rho_inset}
\end{figure}
%----------------------------------------------------------------------------//

Figure \ref{fig:rho_inset} shows the gas density in Snapshot \snapshotB{} for the inner 25\rg{} where the accretion disk has roughly reached inflow equilibrium, and the small inset plot shows the the full simulation grid out to 500\rg{}.  The full simulation grid includes the geometrically thick gas torus extending from $\sim100$\rg{} to $\sim300$\rg{}.  The densities in the funnel regions are several orders of magnitude lower than the densities in the optically thick accretion disk and gas torus.  We discuss the implications of the low density funnel region and the impact of the torus geometry in Section \ref{s-results:comparison_a++}.

The net cooling in the \Athena{} RMHD simulations is given by
%------------------------------  Equation  ----------------------------------
\begin{equation}
\label{eq:netcooling}
    \dot{C} = c \rho \left( \kappa_{\rm P} aT_{\rm g}^4 - \kappa_E E_r\right)+c\rho\kappa_{\rm es}\frac{4kT_{\rm g} - \langle h \nu \rangle}{m_e c^2} E_r,
\end{equation}
%----------------------------------------------------------------------------//
where c is the speed of light, $\rho$ is the gas density, $\kappa_{\rm P}$ is the Planck mean opacity, $a$ is the Planck temperature constant, $T_{\rm g}$ is the gas temperature, $\kappa_{\rm E}$ is the energy mean opacity, $E_{\rm r}$ is the radiation energy density, $\kappa_{\rm es}$ is the electron scattering opacity, $k$ is the Boltzmann constant, $h$ is the Planck constant, $\langle h\nu \rangle$ is the average photon energy, and $m_e$ is the electron mass.  The first term is the frequency and angle integrated free-free emissivity $\eta_{\rm ff} = c\rho\kappa_{\rm P}aT_{\rm g}^{4}$.  The second term is the heating term associated with absorption, and the last term is the net Compton cooling. In the RMHD simulations, the radiation field is assumed to be blackbody so $\langle h\nu \rangle = 4kT_{\rm r}$, where $T_{\rm r}$ is the radiation temperature $T_{\rm r} = (E_{\rm r}/a)^{1/4}$.  The simulations also assume that $\kappa_{\rm E} = \kappa_P$.  These assumptions and their impact on the gas temperature distribution and the spectra that result will be discussed further in Section \ref{s-results}.

%------------------------------  Figure  ----------------------------------
\begin{figure}[t!]
    \centering
    \includegraphics[width=\columnwidth]{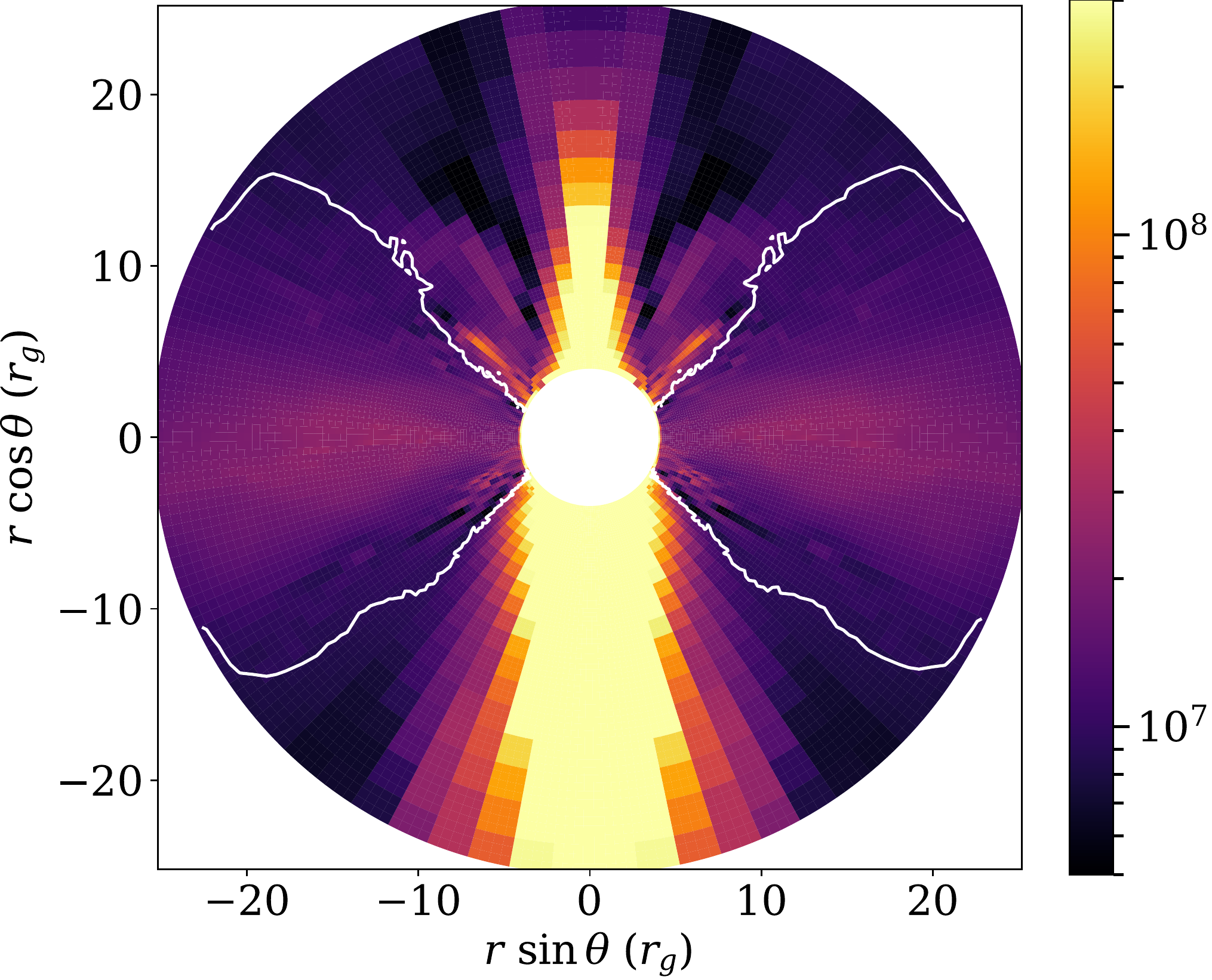}
    \caption{Gas temperature T$_{\rm g}$ (K) shown for Snapshot \snapshotB{}.  A white contour line corresponds to $F_{\rm r}/cE_{\rm r} = 0.3$, which is roughly equivalent to the effective photosphere boundary and defines the polar funnel angle $\theta_{\rm f} = 50^{\circ}$ from the polar axis. The apparent anisotropy of the gas temperature in the funnel regions above and below the disk is a result of this particular simulation taken at this moment in time.  The temperatures were capped at a maximum of $3\times10^8$ K and a minimum of $10^6$ K.}
    \label{fig:tgas18000}
\end{figure}
%--------------------------------------------------------------------------//

The gas temperature of the same snapshot in Figure \ref{fig:rho_inset} is shown in Figure \ref{fig:tgas18000}.  Note that the apparent asymmetry of the gas temperature in the funnel regions above and below the disk is due to the randomness in the flow at the time this snapshot was taken. Prior to post-processing, we set a lower limit on the gas temperature of $10^6$ K and an upper limit of $3\times10^8$ K (except for the multi-group snapshots, which we set the upper limit to $10^9$ K).  The gas temperature is hottest in the funnel region where it hits the temperature cap of $3\times10^{8}$ K, and the gas in the accretion disk peaks at a few$\times10^7$ K.  Although the temperatures in the funnel regions are large, the corresponding gas densities from Figure \ref{fig:rho_inset} are small ($10^{-8}$ g $\rm{cm}^{-2}$) so the contribution to the emission from the hottest simulation cells is relatively weak.  The white contour lines roughly define the effective photosphere boundary between the accretion disk and the funnel region, defined by $F_{\rm r} / cE_{\rm r} = 0.3$, where $F_{\rm r}$ is the $r\phi$ component of the radiative flux, and $E_{\rm r}$ is the radiation energy density.  This flux ratio is consistent with methods that define the photosphere by integrating to an optical depth $\tau=1$ surface \citep{chandrasekhar1960,kinchetal2019}.  The polar angle of this boundary is used to approximate the funnel opening angle $\theta_{\rm f}$ which is then used to calculate the luminosity, spectra, and images in Section \ref{s-results:spectra_imaging}.

%------------------------------  Figure  ----------------------------------
\begin{figure*}[ht!]
    \centering
    \includegraphics[width=0.95\textwidth]{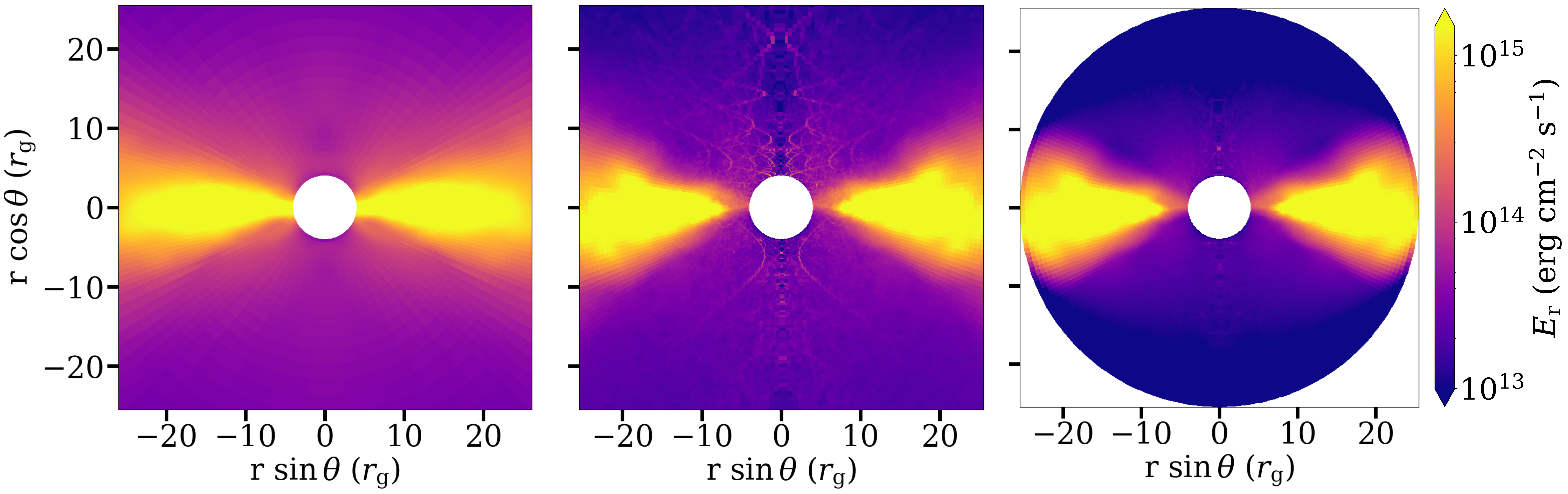}
    \caption{Azimuthally-averaged radiation energy density of Snapshot \snapshotB{} for the \Athena{} RMHD simulation (left panel), the MC calculation (middle panel), and the MC calculation with the simulation grid truncated at $r = 25 r_{\rm g}$ (right panel).  In the left and middle panels, only the inner $25 r_{\rm g}$ are plotted here for comparison, but the full simulation grids extend out to 500\rg{}.}
    \label{fig:ermc_compare}
\end{figure*}
%----------------------------------------------------------------------------//

%**********************************************************************************
%*************************    SPECTRAL CALCULATIONS    ****************************
%**********************************************************************************
\subsection{Spectral post-processing}
\label{s-methods:spectral}
Here we describe the methods used for performing our spectral analysis.  Spectra were generated for each azimuthally averaged snapshot, truncating the calculation to only model MC transfer within 25\rg{}.  The properties of all photon samples leaving the domain at 25 \rg{} are tabulated in a list, which is then used to generate spectra.  This truncation radius was chosen primarily because the outer disk radii are not yet in steady state. In particular, the initial torus is thick, which requires an extremely large radiation pressure.  Hence, this torus is not in thermal equilibrium, and is rapidly cooling. We then ran the MC code on a copy of each snapshot grid, initializing $10^7$ photons for Snapshots \snapshotAA{} and \snapshotAB{}, and about $10^8$ photons for the other two snapshots.  The reason for the difference is due to the larger optical depths and mass accretion rates in \snapshotAA{} and \snapshotAB{} that result in a factor of $\sim10$ difference in number of scatterings per photon sample.  Recall that the number of scatterings per photon sample is proportional to the square of the optical depth.  Increasing the number of photons in the MC calculations greatly improves the counting statistics, however we found that more than $10^7$ photons for those snapshots became too computationally expensive due to the large scattering optical depths.

Photons that escaped the $25$\rg{} simulation domain were collected and distributed into 64 photon energy bins ranging from 0.1 keV to 60 keV, and eight direction angles. By direction angle, we mean the angle $\theta_{\rm p}$ that the photon momentum vector makes with the polar axis.  We use the subscript \textit{$\rm p$} to distinguish angles related to the photon momentum from those related to spherical polar coordinate angles. For example,
\begin{equation}
\theta_{\rm p} = \arccos \left(\frac{\mathbf{p} \cdot \hat{z}}{|\mathbf{p}|}\right),
\end{equation}
where $\mathbf{p}$ is the photon momentum vector. These angle bins are distributed uniformly in $\cos \theta_{\rm p}$ and integrated over azimuthal direction angle $\phi_{\rm p}$.  When binning, we do not distinguish between photons leaving above or below the disk. For example, photons with $\theta_{\rm p} \sim 0$ will be placed in the same bin as photons with $\theta_{\rm p} \sim \pi$.

We select only photons which escape through a ``funnel'' like region above and below the disk.  Specifically, we only bin photons within a coordinate opening angle of $\theta_{\rm f}$ from the polar axes, retaining photons leaving the domain at $\theta < \theta_{\rm f}$ or $\theta > \frac{\pi}{2} - \theta_{\rm f}$. This excludes photon samples that leave domain closer to the midplane. Such photons would almost certainly be further scattered in the optically thick accretion disk if we extended our domain outwards.  Hence, we select our funnel opening angle $\theta_{\rm f}$ to roughly correspond to the location of the disk photosphere at 25\rg{}.  The approximate values for this funnel opening angle are listed for each snapshot in Table \ref{tab:parameters}.  

Due to these selections, the resulting spectra are only expected to be useful estimates of the hard X-ray emission as the softer X-ray will have a significant contribution from regions with $r > 25 r_{\rm g}$.  We also cannot infer much about the angular distribution of the escaping photons for angles that are more edge-on than $\theta_{\rm f}$ as such photons would likely interact with an optically thick flow beyond $r = 25 r_{\rm g}$.

In the case of snapshot \snapshotB{} we also perform a MC calculation using the full simulation domain.  In this case we collect all photon samples leaving the domain, but find that the spectrum of the escaping radiation is dominated by contributions from the torus. Due to large optical depths in the outer torus, the calculation is computationally expensive and run with fewer photons, yielding a lower signal-to-noise spectrum. For these reasons, we do not report spectra from these runs, but we do use the cell-averaged radiation outputs for comparison with the truncated runs described above.

%%%%%%%%%%%%%%%%%%%%%%%%%%%%%%%%%%%%%%%%%%%%%%%%%%%%%%%%%%%%%%%%%%%%%%%%%%%%%%%%%%%%%%%%%%%%%
%%%%%%%%%%%%%%%%%%%%%%%%%%%%%%%%%%%     RESULTS    %%%%%%%%%%%%%%%%%%%%%%%%%%%%%%%%%%%%%%%%%% %%%%%%%%%%%%%%%%%%%%%%%%%%%%%%%%%%%%%%%%%%%%%%%%%%%%%%%%%%%%%%%%%%%%%%%%%%%%%%%%%%%%%%%%%%%%%

%------------------------------  Figure  ----------------------------------
\begin{figure}[ht!]
    \centering
    \includegraphics[width=1.0\columnwidth]{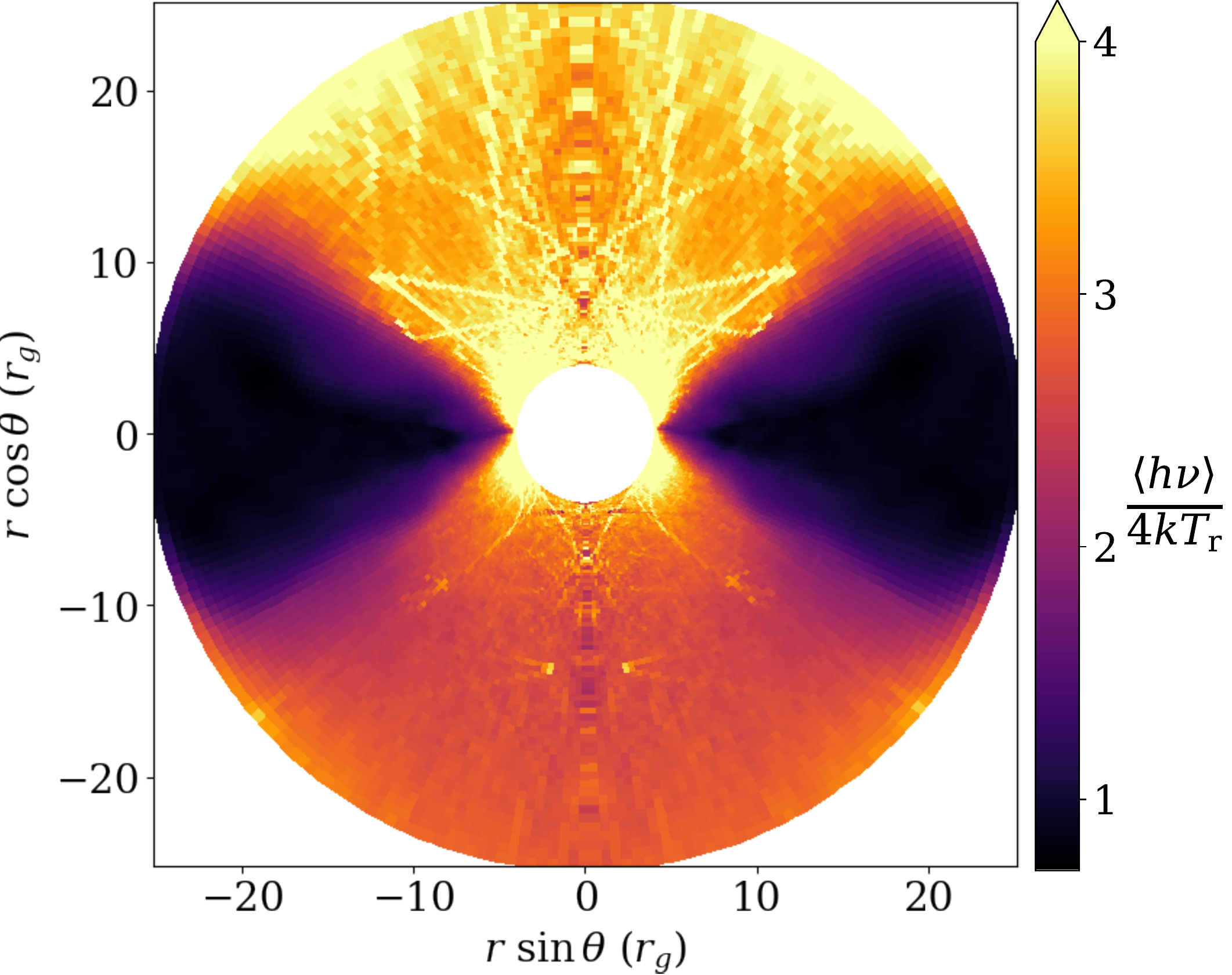}
    \caption{Ratio of the mean photon radiation energy $\langle h\nu \rangle$ calculated in the Monte Carlo code and the radiation energy $4kT_{\rm r}$ in the RMHD simulation for Snapshot \snapshotB{} where $T_{\rm r}$ is the radiation temperature (assuming the blackbody approximation).  The streaks in the funnel region are artifacts of low photon statistics in the Monte Carlo calculation.}
    \label{fig:hnu4kTr}
\end{figure}
%----------------------------------------------------------------------------//

\section{Results}
\label{s-results}
%------------------------------  Figure  ----------------------------------
\begin{figure*}[ht!]
    \centering
    \plotone{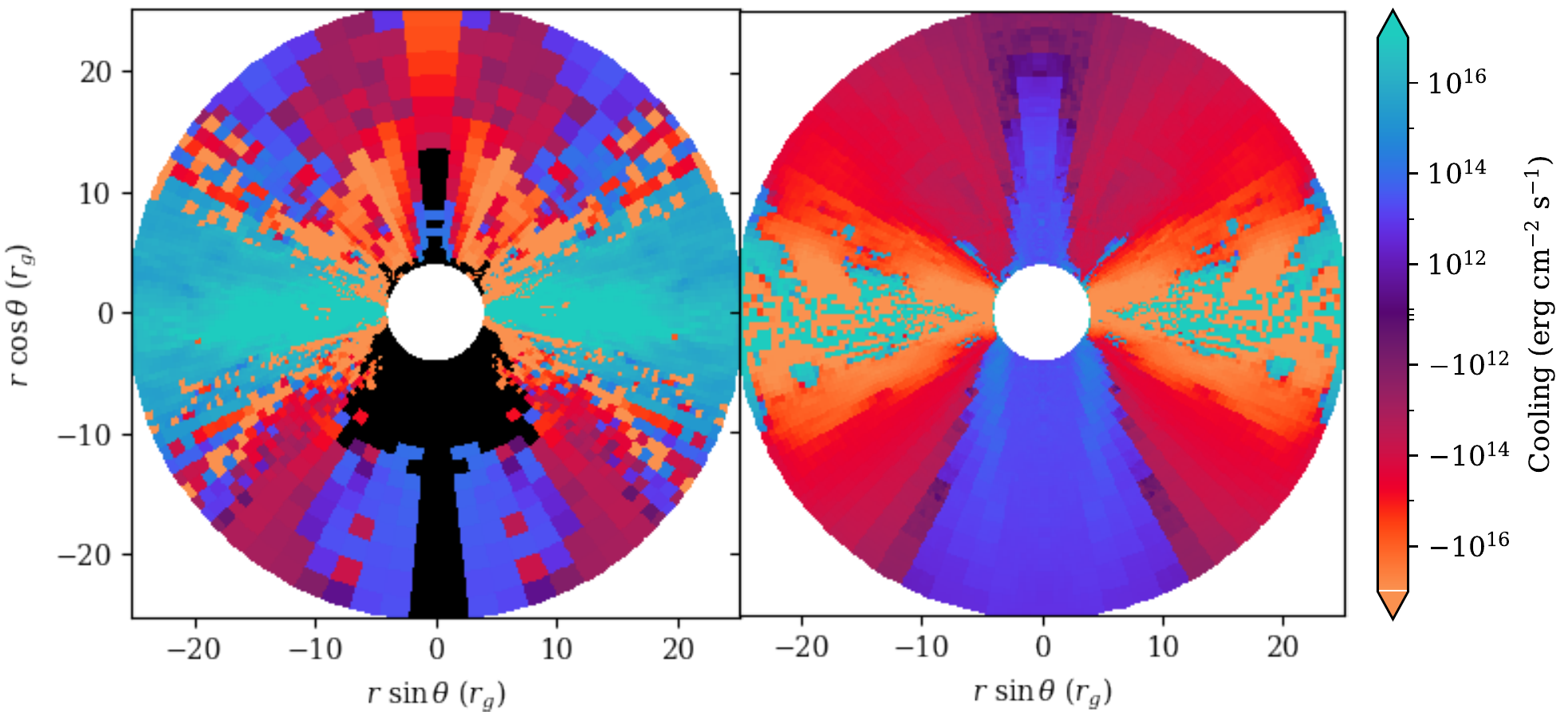}
    \caption{Comparison of the net cooling of Snapshot \snapshotB{} in the \Athena{} RMHD simulation (left panel) and the Monte Carlo $25 r_{\rm g}$ calculation (right panel).  The snapshot has been azimuthally averaged in both cases.  The net cooling is given by Equation \ref{eq:netcooling} where positive values signify net cooling and negative values imply net heating.  The black cells extending into the photosphere in the RMHD calculation are artifacts of this particular moment in the simulation.}
    \label{fig:cooling}
\end{figure*}
%----------------------------------------------------------------------------//

%**********************************************************************************
%*************************    MC MOMENT COMPARISON    ****************************
%**********************************************************************************

\subsection{Comparing \Athena{} with Monte Carlo}
\label{s-results:comparison_a++}

We first compare cell averaged quantities.  Figure \ref{fig:ermc_compare} shows a comparison of $E_{\rm r}$ computed with MC to the azimuthally-averaged $E_{\rm r}$ from the \Athena{} simulation snapshot.  This figure is for Snapshot \snapshotB{}, but the result is representative of all four snapshots in this analysis. The RMHD simulation result is plotted in the left panel, and two MC calculations are plotted in the middle and right panels respectively.  The middle panel shows the results of an MC calculation using the full simulation domain out to $\sim500$\rg{}, whereas the right panel shows the MC calculation when the grid is truncated at $25$\rg{}.  The two MC calculations show precise agreement in the accretion disk, where the radiation field is nearly in radiative equilibrium with the gas. They also agree reasonably well in the funnel regions, deviating by only a small factor near the outer edge of the truncated domain. This suggests that the $E_{\rm r}$ in the inner $25$\rg{} is dominated by the locally emitted radiation field, since the truncated calculations have no incoming photons on the boundary.  Therefore, the radiation from the cooling torus, which dominates the overall emission in the full domain calculation, is not providing a significant contribution in the inner disk region.  Our comparison suggests that radiation outside the truncated domain is contributing $\lesssim 30\%$ near 25\rg{}, and $\lesssim 15\%$ near the photosphere boundary of the disk.

Note that at the very edge of the truncation boundary, $E_{\rm r}$ is slightly lower compared to the radiation energy density in the full domain calculation, as the truncated calculation assumes no incoming radiation flux.  The more noticeable streaks of high $E_{\rm r}$ noise in the funnel regions in the MC full domain calculation are attributed to the factor 10 fewer photons used to compute the full grid, resulting in a larger statistical variance.

Comparing the gray RMHD module $E_{\rm r}$ to the MC calculations in Figure \ref{fig:ermc_compare}, they also appear to agree within a factor of order unity in the accretion disk midplane, but start to deviate more significantly as one transitions into the funnel region.  In the funnel, this deviation is as much as a factor 10.  The MC calculations find a significantly lower $E_{\rm r}$ in the funnel region.  This mismatch is even more evident in Figure \ref{fig:hnu4kTr}, which shows the ratio of the two calculated energy quantities: the mean photon energy $\langle h\nu \rangle$ calculated by the truncated MC calculation, and the mean photon energy $4k T_{\rm r}$ assumed in the \Athena{} RMHD module.  The dark regions where the ratio is of order unity show that the MC and \Athena{} generally agree in the accretion disk, but deviate in the funnel region above and below the disk.  In these regions, the MC calculates that average photon energy is at least three times higher than assumed in the RMHD run.  The assumption that the radiation field is approximately blackbody works well for the optically thick accretion disk regions, but is inadequate in the optically thin funnel regions.

In Figure \ref{fig:cooling}, we compare the resulting cooling computed by the \Athena{} RMHD simulation (left panel) to the same term evaluated by the MC calculation (right panel) for the same snapshot.  The net cooling is calculated using Equation \ref{eq:netcooling} where positive values indicate cooling and negative values indicate heating.  Since the Compton cooling is the dominant term in equation~(\ref{eq:netcooling}) within the funnel region, this comparison is strongly dependent on the degree to which $\langle h\nu \rangle$ differs from $4kT_{\rm r}$ and the ratio of $E_{\rm r}$ in the MC calculations relative to RMHD (see Figures \ref{fig:ermc_compare} and \ref{fig:hnu4kTr}).  We find that the cooling calculated by the MC code deviates significantly from that of the RMHD simulation, particularly in the funnel regions, where the MC code shows significantly more heating and less cooling.  In the accretion disk, the MC code provides slightly less cooling than the RMHD simulation does. The amplitude of the cooling is large in this region because $E_{\rm r}$ is large.  Even though the disk is optically thick, it is hot enough that the Compton term dominates over free-free emission and absorption in both the RMHD and MC calculations.  Since $\langle h\nu \rangle \simeq 4 k T_{\rm g}$ to within a few percent, even small statistical noise in the MC calculation will cause either a large net heating or cooling term here.  Hence, most of the fluctuation seen in the MC calculation in the disk is due to noise in the MC calculation.

These results suggest that if the RMHD simulations had a better estimate for $\langle h\nu \rangle$ it would be higher than $4kT_{\rm r}$.  In an approximate steady-state with the Compton cooling term dominating in the funnel region, one expects the $\langle h\nu \rangle \simeq 4kT_{\rm g}$.  By underestimating $\langle h\nu \rangle$, the RMHD simulations tend to underestimate $T_{\rm g}$ in the optically thin regions above the disk and near the photosphere.  This underestimate also tends to increase $T_{\rm r}$ to better balance $T_{\rm g}$, causing the RMHD simulations to overestimate $E_{\rm r}$, consistent with our findings above.  This also means that spectra computed from these snapshots will have lower average photon energies than one might obtain in a simulation with more self-consistent thermodynamics, which would yield higher $T_{\rm g}$ and harder X-ray spectra.  We explore the implications of this in Section~\ref{s-results:multigroup}.

%**********************************************************************************
%***************************    SPECTRA & IMAGING    ******************************
%**********************************************************************************
\subsection{Post-processed spectra and Compton Cooling from Gray RMHD simulations}
\label{s-results:spectra_imaging}

\begin{figure}[ht!]
    \centering
    \includegraphics[width=\columnwidth]{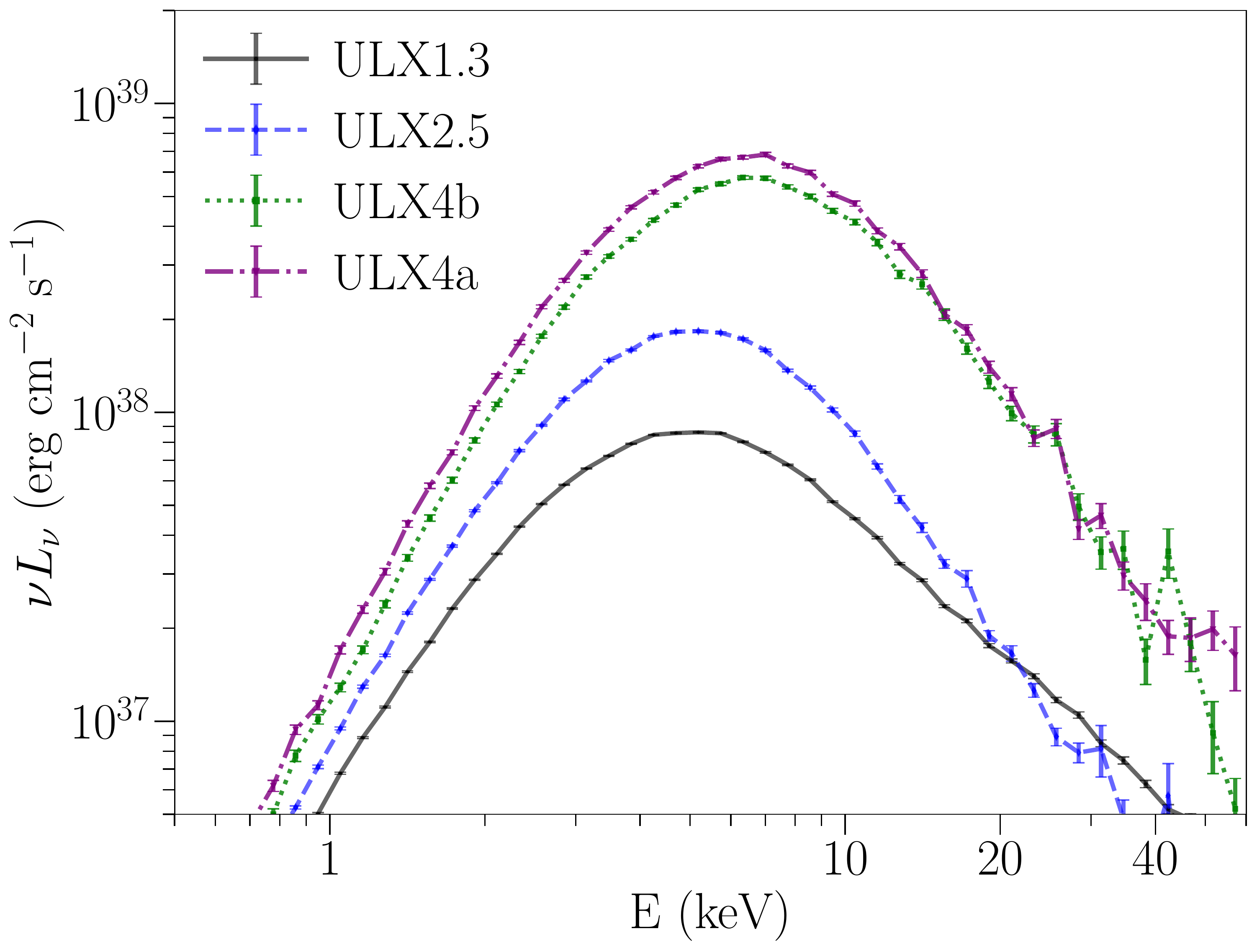}
    \caption{Monte Carlo post-processed X-ray spectra from the gray RMHD simulation snapshots.  From top to bottom: \snapshotAA{} (magenta dash-dot line), \snapshotAB{} (green dotted line), \snapshotB{} (blue dashed line), and \snapshotC{} (black solid line).  Snapshots \snapshotAA{} and \snapshotAB{} were taken from the same simulation run, while \snapshotB{} and \snapshotC{} are both from independent simulations.  Note that these spectra only include the inner $25$\rg{} emission escaping out through a polar funnel angle $\theta_{\rm f}$ specified in Table \ref{tab:parameters} for each snapshot.}
    \label{fig:multispecs}
\end{figure}

We present post-processed X-ray spectra for the four gray RMHD snapshots in Figure \ref{fig:multispecs}. For the lower $\dot{M}$ snapshots (\snapshotC{} and \snapshotB{}), the spectral peaks are roughly around $5$ keV, whereas for the higher $\dot{M}$ snapshots (\snapshotAA{} and \snapshotAB{}) have peaks that are shifted slightly to around $7$ keV. The hard X-ray tails appear to follow power-laws, which we characterize with \texttt{XSPEC} \citep{arnaud1996} model fits in Section \ref{s-results:models}. Due to the truncation of the simulation grids at 25\rg{} prior to post-processing, the softer X-ray emission that should be coming from larger radii is largely absent in these spectra so only the hard X-ray are self-consistently modeled.  

%------------------------------  Figure  ----------------------------------
%--------------------------------------------------------------------------//

The frequency-integrated luminosities for each spectrum are tabulated in Table \ref{tab:parameters}.  We label these funnel luminosities $L_{\rm f}$ to emphasize that we only tabulate the contributions from photons leaving the domain at coordinate $\theta$ within an angle $\theta_{\rm f}$ of the polar axes.  Note that these are mostly hard X-ray luminosities due to the missing soft emission from the outer disk.  The contour lines in Figure \ref{fig:tgas18000} approximate the funnel opening angle for \snapshotB{} ($\theta_{\rm f}=50^{\circ}$) which we show as a representative snapshot.  The $\theta_{\rm f}$ and corresponding funnel luminosities $L_{\rm f}$ are listed for each snapshot in Table \ref{tab:parameters}.

Photons emerging closer to the disk midplane than $\theta_{\rm f}$ are excluded because their escape from the truncated domain at $r=25 r_{\rm g}$ is largely artificial. If we had instead extended our MC calculation domain outward in radius, these photons would likely experience additional scattering and absorption in the optical thick flow before escaping. Slight variations in $\theta_{\rm f}$ can have a modest effect on the funnel luminosity and the resulting spectral shape.  For example, in the case of \snapshotB{} the funnel luminosity varied by less than $17\%$ when varying $\theta_{\rm f}$ by $\pm 10^{\circ}$.  Choosing a narrower funnel angle ($\theta_{\rm f} = 40^{\circ}$) gave a luminosity of $L_{\rm f} = 2.04\times10^{38}$ erg/s, whereas choosing a wider funnel ($\theta_{\rm f} = 60^{\circ}$) gave a slightly higher luminosity of $L_{\rm f}=3.36\times10^{38}$ erg/s. Increasing $\theta_{\rm f}$, however, results in a slight softening of the spectrum as there is an increase in the flux of photons escaping below what would be the photosphere in a more extended domain.  These photons tend to be softer because they are emitted from the cooler regions of the disk. We found this to be true for all snapshot spectra in this analysis.

The radiative efficiency calculated for Snapshot \snapshotB{} is $\eta_{\rm f}=1.51\%$ for a nominal mass accretion rate of $-2.53\dot{M}_{\rm Edd}.$\footnote{We define $\dot{M}_{\rm Edd}$ assuming 10\% efficiency, but $\dot{M}$ itself is independent of our assumed efficiency.}  We report the calculated $\eta_{\rm f}$ for each snapshot in Table~\ref{tab:parameters}.  Generally for super-Eddington accretion, it is expected that the radiative efficiency will be lower than the $\sim5-10\%$ inferred for thin disks, decreasing as accretion rate increases.  We do generally find lower efficiencies, but the results are not completely consistent with expectations. Comparing the efficiencies for snapshots \snapshotB{} and \snapshotC{} which have similar $\theta_{\rm f}$, we infer a slightly lower efficiency as accretion rate increases.  Snapshots \snapshotAA{} and \snapshotAB{}, however, have higher $\dot{M}$ than the other snapshots, but also show a higher $\eta_{\rm f}$.  It is possible that this deviation from the expected trend is a result of our truncation of the calculation at $r = 25 r_{\rm g}$ and merits more consideration in future work exploring a wider range of $\dot{M}$.

\begin{figure}[t]
    \centering
    \includegraphics[width=\columnwidth]{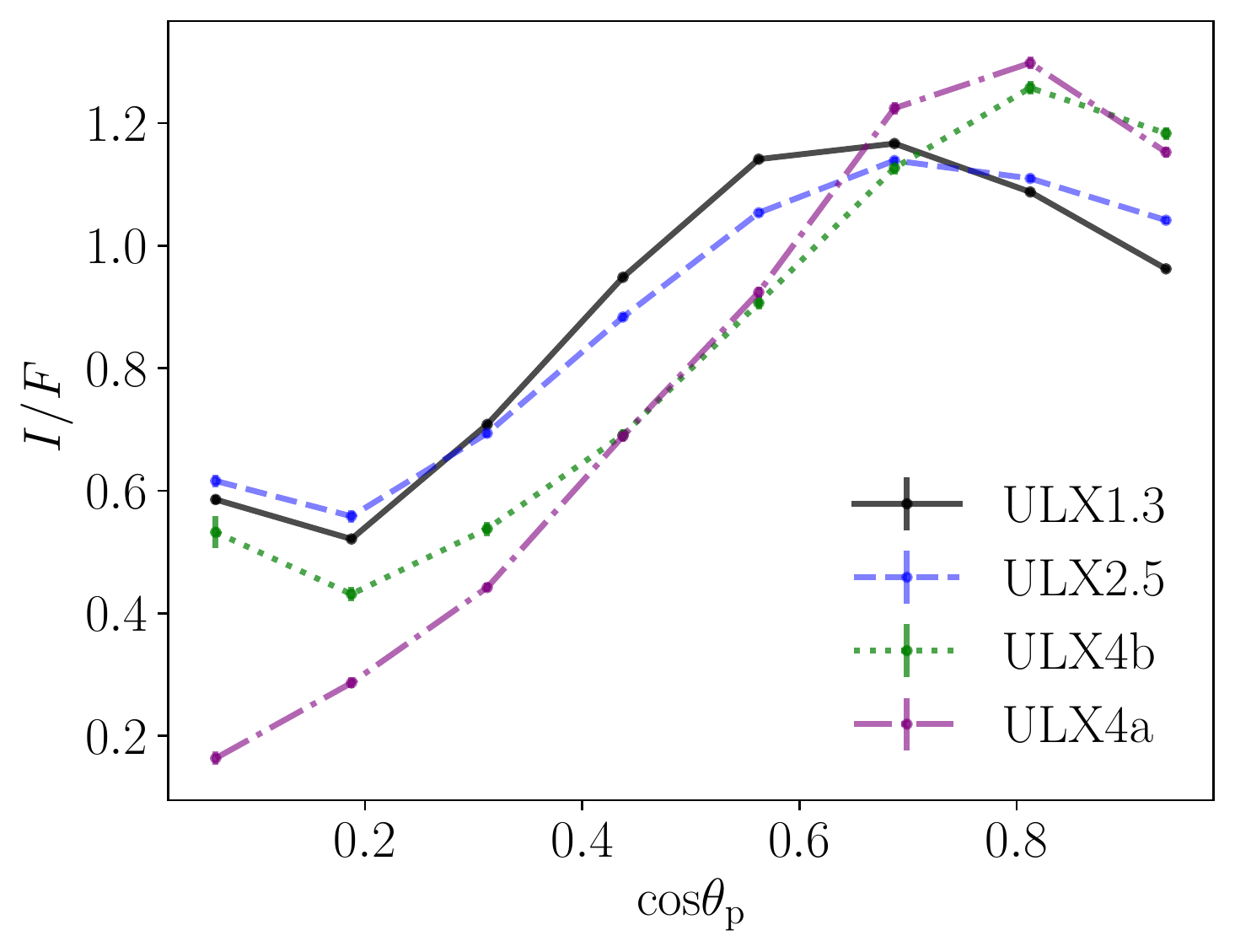}
    \caption{Flux fractions of the specific intensity $I$ to the isotropic flux $F$ as a function of inclination angle in terms of cos$\theta$ for Snapshots \snapshotAA{} (pink dash-dotted line), \snapshotAB{} (green dotted line), \snapshotB{} (blue dashed line), and \snapshotC{} (black solid line).  Note that each snapshot spectrum was generated using only photons which escape through a funnel opening of polar angle $\theta_{\rm f}$ specified in Table \ref{tab:parameters}.  Face-on viewing corresponds to $\cos\theta_{\rm{p}}=1$ and edge-on viewing corresponds to $\cos\theta_{\rm{p}}=0$.}
    \label{fig:fluxfracs}
\end{figure}

We also examine the angular distribution of the emission, which we  model as the flux fraction (ratio of specific intensity $I$ to the flux $F$) for each spectrum in Figure \ref{fig:fluxfracs}. For observations, this should roughly correspond to the inclination viewing angle dependence with respect to the polar axis. A face-on view of the emission corresponds to $\cos\theta_{\rm p }=1$, and an edge-on view corresponds to $\cos\theta_{\rm p}=0$.  Snapshot \snapshotC{} is shown as the solid black line, \snapshotB{} is the dashed blue line, \snapshotAB{} is the dotted green line, and \snapshotAA{} is the dash-dot pink line.  The flux fraction has been integrated over all frequencies $\nu$ for improved statistics, so the resulting distribution is most representative of the angular distribution near the spectral peak.

Although we show the full distribution for $\cos \theta_{\rm p}$ ranging from 0 to 1, we emphasize that the $\cos \theta_{\rm f}$ ranges from 0.57 for \snapshotC{} to 0.8 for snapshots \snapshotAA{} and \snapshotAB{}. Hence, only the bins with $\cos \theta_{\rm p}$ greater than these values are likely to be well-characterized.  Over this limited range, the angular distributions are relatively flat, but notably do not peak at the most face on inclination bin.  This is contrary to standard expectations where a face on view provides the largest projected area and, thus, the largest flux.  As $\theta_{\rm p}$ approaches the edge-on view the intensity declines by factors of several.  However, we emphasize that the intensity distribution at these angles will undoubtedly be impacted by the extension of the optically thick disk outside of the calculations domain.  For example, the slight rise in the most edge-on bin is almost certainly a result of our artificial truncation of the simulation domain.  Hence, our current results cannot provide reliable predictions about geometric beaming factors.

%------------------------------  Figure  ----------------------------------
%--------------------------------------------------------------------------//

\begin{figure}[ht!]
    \centering
    \includegraphics[width=\columnwidth]{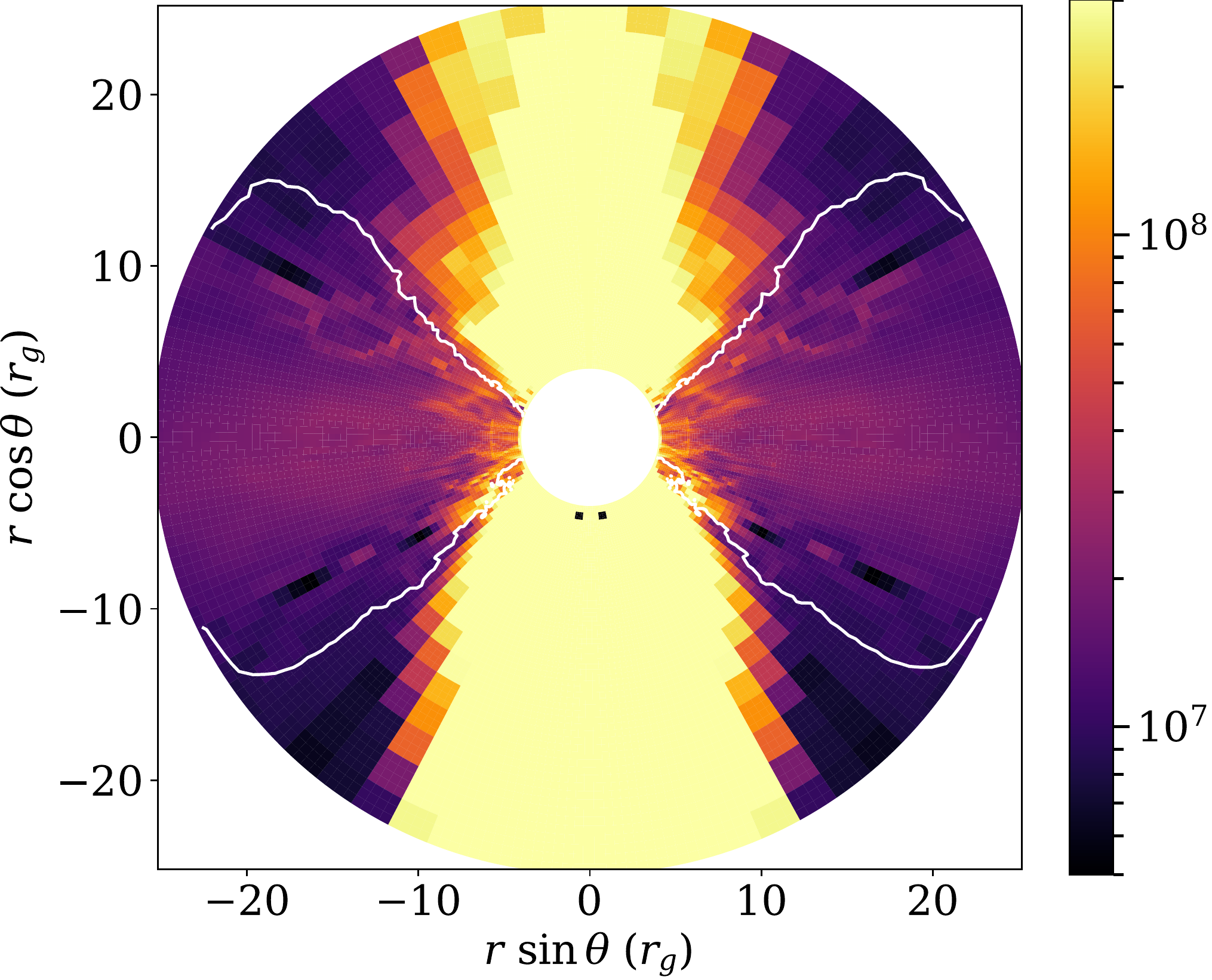}
    \caption{The same as Figure \ref{fig:tgas18000}, except for the multi-group snapshot \snapshotBmulti{} with a temperature upper limit of $T_{\rm g} = 10^9$ K, although the colorbar in this plot goes to $3 \times 10^8$ K for comparison.}
    \label{fig:tgas_18056}
\end{figure}

\begin{figure*}[ht!]
    \centering
    \epsscale{0.8}
     \plotone{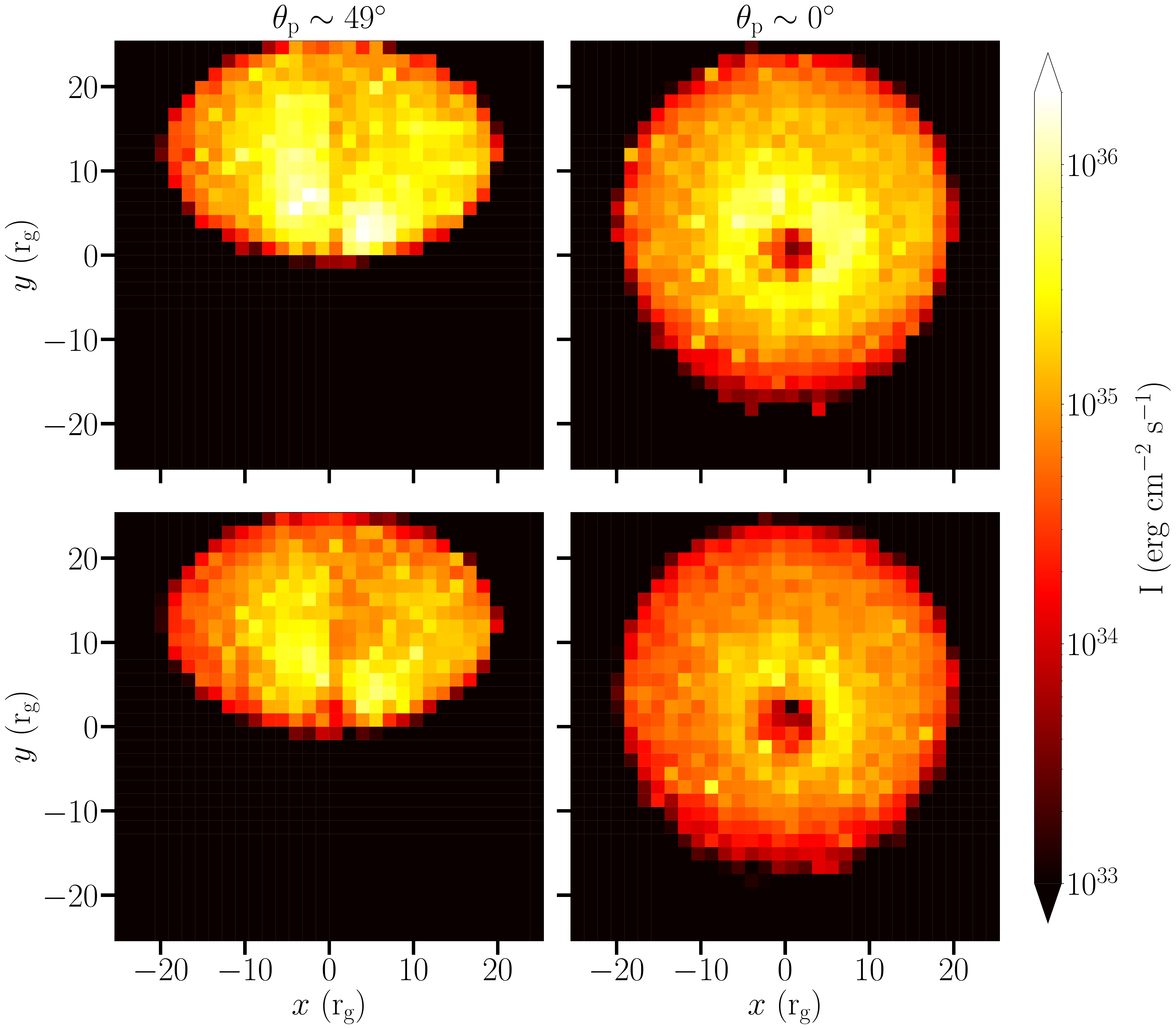}
    \caption{A series of frequency-integrated images showing the emergent radiation intensity for snapshots \snapshotBmulti{} (top row) and \snapshotB{} (bottom row) for two inclination viewing angles.  The left column shows a viewpoint from an inclination of $\theta_{\rm{p}} \sim 49^{\circ}$, which is at the edge of the funnel for these snapshots ($\theta_{\rm f}=50^{\circ}$).  The right column views the face-on inclination ($\theta_{\rm f} = 0 ^{\circ}$).  The photons leaving the simulation domain at $25$ \rg{} were extrapolated out to a distance of 250,000 \rg{} from the black hole.}
    \label{fig:images}
\end{figure*}

To better interpret these results, we show a set of reconstructed images in Figure~\ref{fig:images}, which shows the frequency-integrated intensity from the funnel region at different inclination angles $\theta_{\rm{p}}\sim49^{\circ}$ (left column; funnel edge view) and $\theta_{\rm{p}}=0^{\circ}$ (right column; face-on view) for two snapshots: the gray RMHD snapshot \snapshotB{} (bottom row), and the multi-group RMHD snapshot \snapshotBmulti{} (top row).  We discuss the latter snapshot in detail in the next section.  The corresponding opening angle for both snapshots is $\theta_{\rm f}=50^{\circ}$.  Photons escaping the funnel were extrapolated out to a distance of $\sim250,000$ \rg{} to form these images.  In the face-on case, we see a deficit for photons near the polar axis, along the line of sight to the black hole.  In this region the densities are so low that relatively few photons are scattered or emitted toward the observer.  Near the edge of the funnel, the intensity of the emission appears to brighten compared to the face-on inclination.  This enhancement is consistent with modest amounts of relativistic beaming in the mildly relativistic outflowing gas.  This beaming is largest at these moderate inclinations where both the line-of-sight outflow velocities and scattering optical depths are large.

%------------------------------  Figure  ----------------------------------
%--------------------------------------------------------------------------//
\subsection{A multi-group RMHD approach}
\label{s-results:multigroup}

As discussed in Section~\ref{s-results:comparison_a++}, the $\langle h\nu \rangle = 4kT_{\rm{r}}$ assumption in the gray RMHD simulations likely results in gas temperatures being underestimated in the regions above the optically thick disk. This, in turn, means that the MC spectra we compute are probably softer than they should be if the temperatures were computed with a more self-consistent treatment of Compton scattering. In an effort to recompute the gas temperature, we first tried to use the MC code to calculate the net cooling everywhere in the simulation and balanced this with the dissipation from the RMHD snapshots.  We found, however, that the recomputed temperatures in the funnel had too large of a variance due to the limited photon statistics, and did not consistently converge after several iterations.

Instead, we utilized the multi-group radiation module described in \cite{jiang2022}, which extends the gray radiation scheme in \cite{jiang2021} to include frequency dependence and treats Compton scattering using a Kompaneets-like approximation for the electron scattering source term. We used 20 logarithmically distributed frequency groups to cover the frequency space over three orders of magnitude, which increased the computational cost by a similar factor.  Hence, a full three-dimensional simulation with this method would be extremely computationally expensive.  Here, we instead begin the multi-group simulation by assuming the initial spectrum to be blackbody, and thus restart the gray simulation and run for a time just long enough for the gas above the disk to reach a new temperature equilibrium. This makes the computational expense feasible for this study because the thermal timescale in the funnel region is very short. 

We performed the multi-group procedure for two of the four snapshots, \snapshotB{} and \snapshotAA{}, using the restart files from the gray RMHD simulations and running them with 20 frequency groups.  We label the new snapshots from these multi-group runs \snapshotBmulti{} and \snapshotAAmulti{}.  These snapshots were computed at approximately the same time as their gray counterparts, and thus have the same average mass accretion rates (see Table \ref{tab:parameters}) and the density distributions are quite similar.  But, as expected from the MC calculations of cooling rates in the gray snapshots, these new runs find larger gas temperatures in the regions near or above the photosphere of the accretion flow, as shown in Figure \ref{fig:tgas_18056} (compared to its gray counterpart temperature in Figure \ref{fig:tgas18000}). We post-process these snapshots following the same procedures as we did for the gray snapshots.

We do not show the angular distributions of the emitted spectra for these multi-group snapshot calculations because they are rather similar to their counterparts shown in Figure \ref{fig:fluxfracs}. We show, however, a second set of reconstructed images in the top panel of Figure \ref{fig:images} for \snapshotBmulti{}.  Compared to its gray counterpart in the top panel, the overall intensities in the multi-group approach are larger due to the larger temperatures, but the mild relativistic beaming again enhances the intensities for off-axis viewing angles relative to those of the most face-on image.

\begin{figure}[t!]
    \centering
    \includegraphics[width=\columnwidth]{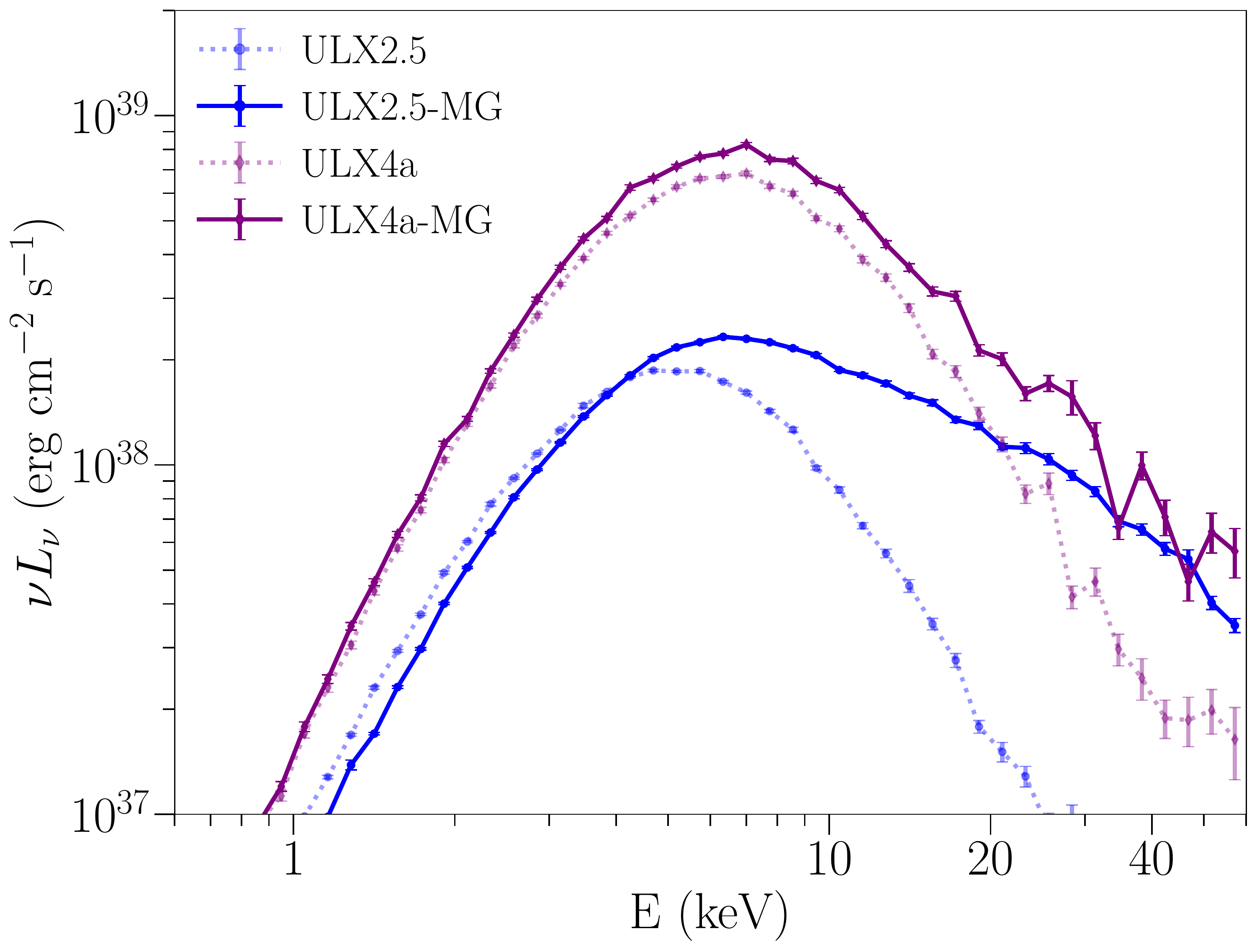}
    \caption{MC post-processed spectra from the gray RMHD simulation snapshots (\snapshotB{} and \snapshotAA{}) shown as the lighter blue and purple colored dotted lines, respectively, and the MC spectra from the multi-group RMHD implementation shown as the corresponding darker solid lines. The spectra for \snapshotB{} and \snapshotBmulti{} were computed for a funnel region of $\theta_{\rm f}=50^{\circ}$, while \snapshotAA{} and 
    \snapshotAAmulti{} were computed for $\theta_{\rm f}=37^{\circ}$.}
    \label{fig:multigroup}
\end{figure}

Figure \ref{fig:multigroup} shows the MC post-processed spectra from the multi-group snapshots compared to their corresponding gray snapshot counterparts.  The multi-group approach leads to harder spectra due to the larger gas temperatures.  This is seen as both a shift in the spectral peak and a somewhat flatter power-law dependence at higher energies. The effect is larger for \snapshotBmulti{} than \snapshotAAmulti{}. The overall luminosity of the funnel for the multi-group spectra are also larger, with $L_{\rm f}=4.73\times10^{38}\ \rm{erg}\ \rm{cm}^{-2}\ \rm{s}^{-1}$ for \snapshotBmulti{}, and $L_{\rm f}=1.31\times10^{39}\ \rm{erg}\ \rm{cm}^{-2}\ \rm{s}^{-1}$ for \snapshotAAmulti{}. There is also a commensurate increase in the radiation efficiencies since the accretion rates were essentially unchanged.

%------------------------------  Figure  ----------------------------------
\begin{figure*}[t]
    \epsscale{0.9}
    \centering
    \plotone{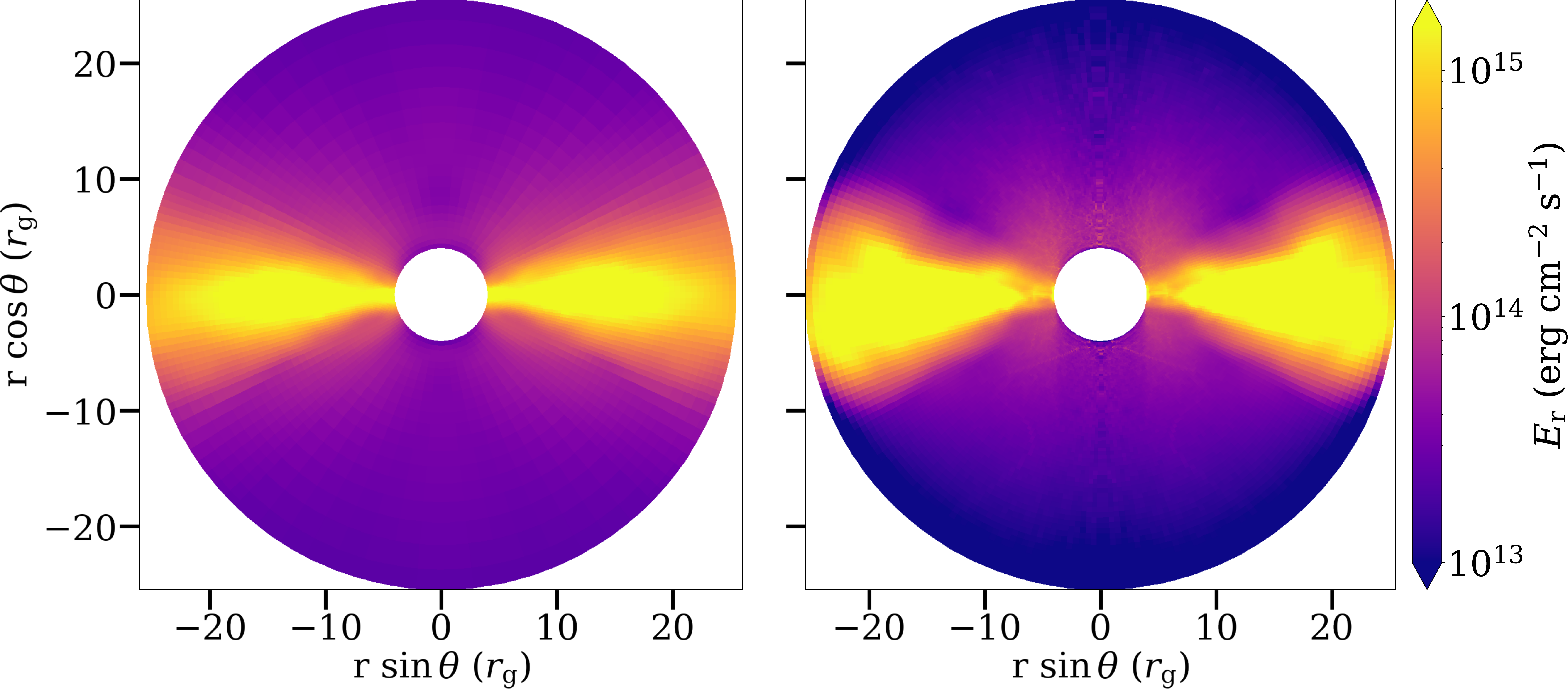}
    \caption{Comparison of the radiation energy density $E_{\rm{r}}$ for \snapshotBmulti{} from the multi-group RMHD snapshot (left panel) and the same quantity calculated by the Monte Carlo module (right panel).  In both cases, the simulation has been azimuthally averaged.}
    \label{fig:multigroup-Er}
\end{figure*}
%----------------------------------------------------------------------------//

Figure \ref{fig:multigroup-Er} shows a comparison of the radiation energy density for \snapshotBmulti{} from the multi-group RMHD simulation (left panel) and the MC calculated radiation energy density (right panel), analogous to the comparison for the gray snapshot \snapshotB{} in left and right panels of Figure \ref{fig:ermc_compare}.  As expected, the $E_{\rm r}$ in the multi-group approach is lower compared to its gray counterpart, and the comparison with MC for the multi-group approach is in closer agreement, although not exact.  Exact agreement is not necessarily expected as the Kompaneets treatment in the RMHD module differs slight from the MC treatment.  It also possible that the (computationally expensive) multi-group calculation has not yet reached full equilibrium.

%**********************************************************************************
%***************************    SPECTRAL MODELS    ********************************
%**********************************************************************************
\subsection{Simulated spectra in comparison with phenomenological models}
\label{s-results:models}

Here we quantitatively characterize the post-processed spectra by utilizing X-ray spectral fitting models commonly used to describe observations of black hole sources.  The motivation here is to get a sense of how the combinations of phenomenological models describe the hard X-ray emission and quantitatively compare between the post-processed gray RMHD spectra and the multi-group RMHD spectra. Although we utilize spectral fitting methodology as a tool to compare our simulations to other models, we emphasize that these are not fits to data, and we make choices in accordance with these considerations. 

We use the X-ray spectral fitting package\texttt{XSPEC} \citep{arnaud1996} version 6.26.1 to explore a few different model combinations.   We note that the two-component (soft+hard) phenomenological model combinations we use here are typically used to fit BHXB spectra whilst ULX spectra can also be described by two component models (as shown using variability studies: \citealt{Middleton2015}), where the components refer to regions in the super-Eddington disk, modified by opacity in the wind and anisotropy (\citealt{Poutanen2007}). In addition, ULX spectra sometimes require a third component at higher energies from a pulsing component (an accretion column: \citealt{brightmanetal2016, Walton2018_sample}) which has led to speculation that a generic hard excess compared to thermal models could indicate the presence of a highly magnetised neutron star (\citealt{pintoreetal2017, Walton2018_sample}). By comparison, our spectra only correspond to the innermost regions and so miss a large portion of the soft X-ray emission from outer radii; we therefore use only one soft X-ray and one hard X-ray component to describe our simulated spectra, and focus mainly on the hard X-ray part of the spectra. Since we only seek to characterize our simulated spectra, we do not include any absorption components that are typically used to account for the interstellar medium along the line of sight.  

The simulated spectra were transcribed into table models containing energies and fluxes that could be loaded into\texttt{XSPEC}.  For each simulated model, the energy range was limited to $3$ -- $50$ keV.  We used the \texttt{fakeit none} command to generate an artificial ``dataset'' for each simulated table model.  For the required response file during this process, we input a \textit{NuSTAR} FPMA detector response file provided by one of the \cite{gurpideetal2021} observations of NGC 1313 X-1 (see following subsection).  All artificial datasets were generated assuming a 100 ks exposure time.  The systematic error was set to 5\%, a large fraction compared to the error from the counting statistics. The inclusion of a large systematic error is chosen so that this fitting procedure gives a reasonable characterization of the hard X-ray tails in our synthetic spectrum.  This is, of course, different from standard fitting procedures to data where bins with more counts generally have higher signal-to-noise and are thus weighted more heavily in the fit.  The inclusion of a large systematic error results in the bins in the hard X-ray tail being treated on a more equal footing with those near the peak. If not included, the best-fit spectral slopes are notably flatter than our synthetic spectra in the hard X-ray tail.  This is due to relatively small changes in the fit near the peak which drive larger changes to $\chi^2$ than the large deviations in the X-ray tail.  The procedure employed here, however, provides a reasonable match to both the continuum near the peak and in the X-ray tail.

After the artificial datasets were generated, we chose a sample of model combinations listed in Table \ref{tab:models} along with the corresponding fit parameters.  The model \texttt{bbody} fits a blackbody spectrum with two parameters: a temperature $kT$ (keV) and normalization, the latter given by $N_{\rm{BB}}=L_{39}/D_{10}^2$, where $L_{39}$ is the source luminosity in units of $10^{39}$ and $D_{10}$ is the distance to the source in units of 10 kpc.  Similarly, the \texttt{diskbb} model is a multi-temperature blackbody accretion disk (without a colour temperature correction factor) and has two free parameters: the inner disk temperature $T_{\rm{in}}$ (K) and a normalization parameter defined as $N_{\rm{DBB}}=(R_{\rm{in}}/D_{10})^{2}\cos\theta$ where $R_{\rm{in}}$ is the apparent inner disk radius in km, $D_{10}$ is the distance to the source in units of 10 kpc, and $\theta$ is the disk inclination angle at which $\theta=0$ is face-on \citep{mitsudaetal1984}.

In addition to the accretion disk models, a hard X-ray component was added to characterize the hard X-ray flux.  We chose a power-law model \texttt{pow} which has two free parameters: the power-law index $\Gamma$ (such that flux goes as $E^{-\Gamma}$), and a normalization parameter. The other model we chose to characterize the hard X-ray spectrum is the X-ray Comptonization model \texttt{simpl} \citep{steineretal2009}.  \texttt{simpl} is a convolution model that approximately Compton up-scatters a fraction $f_{\rm sc}$ of seed photons from the \texttt{bbody} or \texttt{diskbb} models.  These up-scattered photons form a hard X-ray power-law tail with index $\Gamma$.  We assume that photons will only be up-scattered, leaving two free fit parameters, similar to \texttt{pow}. For the model combinations that include \texttt{simpl}, we set an upper limit on the scattering fraction of $f_{\rm sc}=60\%$ as this parameter was not well constrained at higher $f_{\rm sc}$.

We report the fit results in Table \ref{tab:models}.  We only report a few significant digits without the errors as these are essentially model fits to simulated data, in contrast to model fits to observed data.  Any errors computed here would strongly depend on the chosen systematic and stochastic errors, and are not physically meaningful. We do not report goodness-of-fit for similar reasons.  

For the gray snapshots, we find a rather steep power-law index of $\Gamma > 4$ is required for all fits when \texttt{simpl} is used.  The index is still steep, but somewhat flatter when \texttt{pow} is used. With the combination of \texttt{diskbb+pow}, the \texttt{pow} model component dominated the fit and could not adequately describe the softer part of the spectrum. This is partly because we are missing a large portion of the soft X-rays from the outer disk in the simulations, but is also related to the lack of an absorption model to attenuate the power-law emission at softer energies.  In contrast, the multi-group snapshots were characterized by much flatter hard X-ray tails (e.g. $\Gamma \lesssim 4$) than their gray counterparts, particularly for \snapshotBmulti{} ($\Gamma \lesssim 3$). These flatter indices are in better agreement with most observed spectra \citep{pintoreetal2017,dage2021,gurpideetal2021}.  We also generally find high scattering fractions (e.g. $f_{\rm sc} \sim 60\%$) for the multi-group spectra compared to their gray counterparts (e.g. $f_{\rm sc} \sim 40--50\%$), particularly for the \texttt{diskbb+simpl} model fits. This is indicative of the fact that the power-law extends from near the spectral peak.

\begin{figure}[t!]
    %\epsscale{1.2}
    \centering
    \includegraphics[width=\columnwidth]{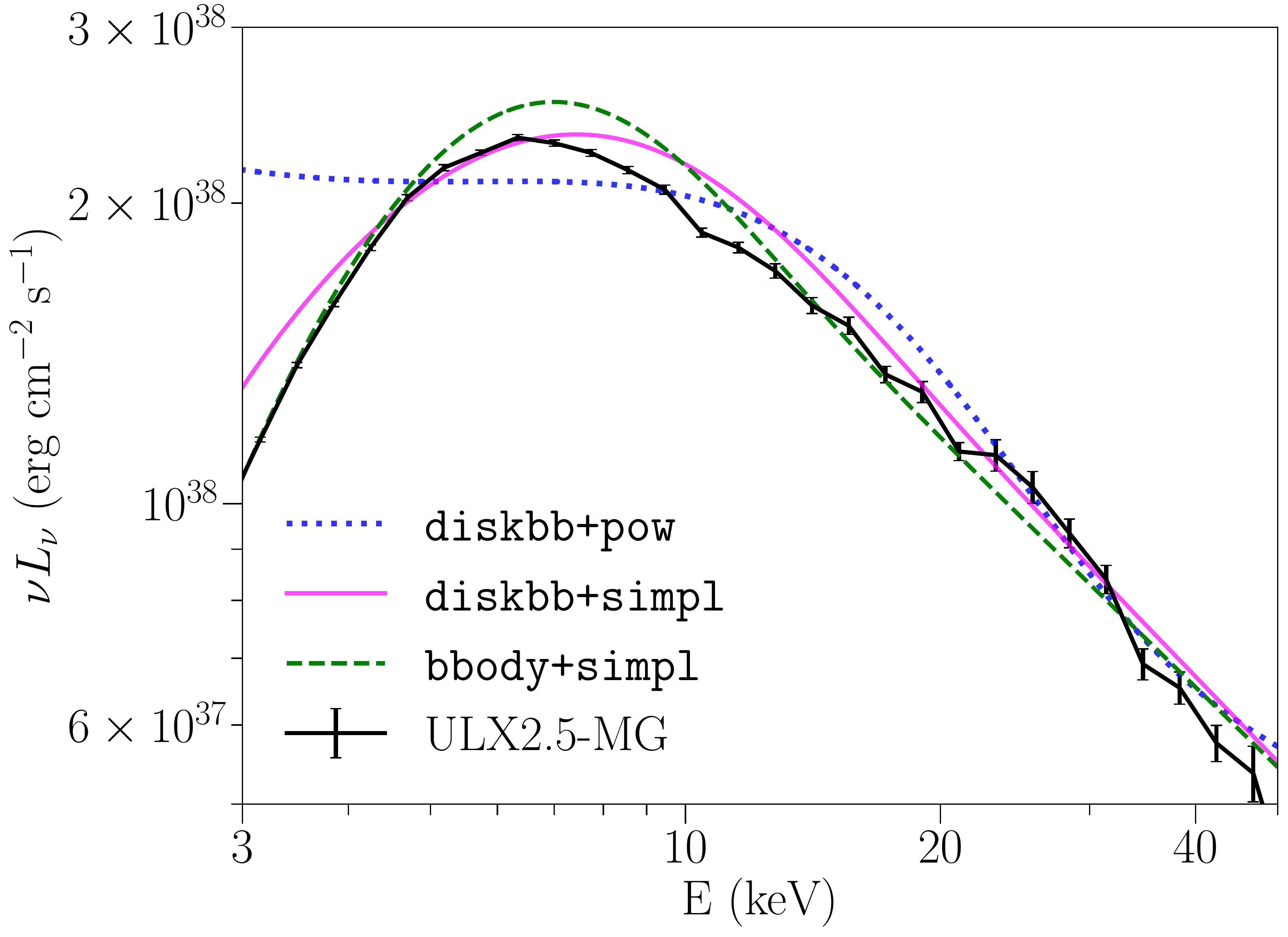}
    \caption{Comparison of three X-ray spectral fitting models to the post-processed multi-group RMHD spectrum \snapshotBmulti{} (shown as the black solid line).  The model combinations include two components, one blackbody (\texttt{bbody}) or multi-temperature blackbody accretion disk \texttt{diskbb} model paired with either a power-law (\texttt{pow}) or a hard X-ray Comptonization model (\texttt{simpl}).  The total combinations are shown as the blue dotted line for \texttt{diskbb+pow}, pink solid line for \texttt{diskbb+simpl}, and green dashed line for \texttt{bbody+simpl}. Note that the \texttt{simpl} model in \texttt{XSPEC} would be written as \texttt{simpl$\times$diskbb}, since it is a convolution model.}
    \label{fig:eemodels}
\end{figure}

Figure \ref{fig:eemodels} shows an example of the three model combination fits to \snapshotBmulti{}.  The \snapshotBmulti{} spectrum is shown as the black solid line.  The total \texttt{diskbb+pow} model corresponds to the blue dotted line, which shows the deviation of the fit at the softer end of the spectrum due to the \texttt{pow} component dominating the fit.  Interestingly, the \texttt{bbody+simpl} model more closely fit the simulated spectra compared to the other two model combinations.  For most of the spectra, the \texttt{diskbb} component in the \texttt{diskbb+simpl} model was slightly broader than the simulated spectrum as seen by the pink solid line in Figure \ref{fig:eemodels}. However, this may again be impacted by the missing soft X-rays from larger radii.

%------------------------------  Table  ----------------------------------
\begin{deluxetable*}{ccccccc}
\tablecaption{Model comparisons}
\tablecolumns{7}
\tablehead{\colhead{Model Component} & \colhead{Parameter} & \colhead{\snapshotB{}}  & \colhead{\snapshotBmulti{}} & \colhead{\snapshotAA{}} & \colhead{\snapshotAAmulti{}}}
\startdata
\texttt{bbody}      & kT                & $1.13$  & $1.42$  & $1.46$  & $1.39$ \\
                    & $N_{\rm{BB}}$     & $0.23$  & $0.32$  & $0.85$  & $0.99$ \\
\texttt{simpl}      & $\Gamma_{\rm{S}}$ & $4.38$  & $2.82$  & $4.34$  & $3.52$ \\
                    & $f_{\rm sc}$      & $60\%$  & $60\%$  & $60\%$  & $60\%$\\
\hline
\texttt{diskbb}     & $T_{\rm in}$      & $1.71$  & $2.27$& $2.47$  & $2.04$ \\
                    & $N_{\rm{DBB}}$    & $149.92$ & $56.42$& $116.21$  & $288.64$\\
\texttt{simpl}      & $\Gamma_{\rm{S}}$ & $4.34$  & $2.88$& $4.14$ & $3.57$\\
                    & $f_{\rm sc}$      & $41.5\%$ & $60\%$ & $52.2\%$ & $60\%$ \\
\hline
\texttt{diskbb}     & $T_{\rm in}$       & $2.35$    & $4.30$ & $3.06$  & $3.34$ \\
                    & $N_{\rm{DBB}}$    & $25.39$ & $2.07$ & $37.65$ & $23.36$ \\
\texttt{pow}        & $\Gamma_{\rm{P}}$ & $3.36$ & $2.40$ & $3.03$  &  $2.80$ \\
\enddata
\tablecomments{Comparison of commonly used spectral fitting models when fit to the MC post-processed gray RMHD snapshot spectra (\snapshotB{} and \snapshotAA{}) and the MC post-processed multi-group RMHD snapshot spectra (\snapshotBmulti{} and \snapshotAAmulti{}).  We only report rough values for each model parameter without errors in an effort to get a sense of the relative spectral shape of the simulated spectra for different model combinations.  The accretion disk models used to fit the softer part of the spectrum (\texttt{diskbb} and \texttt{bbody}) were combined with either the Comptonization model \texttt{simpl} or power-law model \texttt{pow} to fit the hard X-ray fluxes.}
\end{deluxetable*}\label{tab:models}
%--------------------------------------------------------------------------//

%------------------------------  Figure  ----------------------------------
%----------------------------------------------------------------------------//

%------------------------------  Table  ----------------------------------
\begin{deluxetable*}{ccccccc}
\tablecaption{Best-fit parameters to NGC 1313 X-1 data using simulated spectral models}
\tablecolumns{7}
\tablehead{\colhead{Model Component} &  \colhead{Parameter}  & \colhead{Typical model} & \colhead{\snapshotB{}} & \colhead{\snapshotBmulti{}} & \colhead{\snapshotAA{}} & \colhead{\snapshotAAmulti{}}}
\startdata
%\multicolumn{7}{c}{\texttt{XSPEC} notation: \texttt{TBabs}$\times$\texttt{gabs}$\times$(\texttt{diskbb}$+$(\texttt{simpl}$\times$\texttt{diskpbb}))} \\
\hline                                                          %% Typical model            %% ULX2.5                   %% ULX2.5-MG
\texttt{TBabs}          & $n_{\rm{H}}$ ($\rm{cm}^{-2}$)     & $0.27_{-0.03}^{+0.02}$    & $0.26_{-0.02}^{+0.02}$    & $0.16_{-0.02}^{+0.03}$    & $0.15_{-0.02}^{+0.02}$    & $0.19_{-0.03}^{+0.02}$   \\
\texttt{diskbb}         & $T_{\rm{in}}$ (keV)               & $0.27_{-0.03}^{+0.03}$    & $0.27_{-0.01}^{+0.01}$    & $0.72_{-0.03}^{+0.04}$    & $0.68_{-0.02}^{+0.03}$    & $0.72_{-0.03}^{+0.03}$  \\
                        & $N_{\rm{DBB}}$                    & $11.2_{-4.3}^{+7.9}$      & $13.83_{-6.2}^{+6.9}$     &  $0.30_{-0.07}^{+0.07}$   & $0.42_{-0.7}^{+0.08}$     & $0.33_{-0.06}^{+0.06}$ \\
\texttt{MC spectrum} & $N$ ($\times10^{-6}$)           & n/a                       & $100.0_{-1.8}^{+1.8}$     & $74.2_{-2.1}^{+2.1}$      &  ${26.5}_{-0.7}^{+0.6}$   & $22.1_{-0.5}^{+0.6}$ \\
\texttt{diskpbb}        & $T_{\rm{in}}$ (keV)               & $3.22_{-0.46}^{+0.62}$    & n/a                       & n/a                       & n/a                       & n/a  \\
                        & $p$                               & $0.56_{-0.02}^{+0.03}$    & n/a                       & n/a                       & n/a                       & n/a  \\
                        & $N_{\rm{p}}$ ($\times10^{-4}$)  & $4.49_{-2.5}^{+5.6}$      & n/a                       & n/a                       & n/a                       & n/a \\
\texttt{simpl}          & $\Gamma$                          & $2.00_{-2}^{+0.53}$       & n/a                       & n/a                       & n/a                       & n/a  \\
                        & $f_{\rm{sc}}$ ($\%$)              & $11.50_{-6.1}^{+7.8}$     & n/a                       & n/a                       & n/a                       & n/a \\
                        & $\chi^2$/dof                      & $373.60/341$              & $777.71/345$              & $403.37/345$              & $445.72/345$              & $399.05/342$ \\
%%% ULX2.5 c1=1.1.15, c2=1.21
%%% reduced chi^2: ngc:1.09, ulx2.5: 1.85, 2.25, 2.20, 1.32, 1.26, 1.17
\enddata
\tablecomments{Best-fit parameter values from the simulated spectral model fits to the combined \textit{XMM-Newton} and \textit{NuSTAR} data of NGC 1313 X-1.  We also show a comparison ``Typical model'' fit to NGC 1313 X-1, for comparison to the fits from the four post-processed or simulated snapshot spectral models.  The two models \snapshotAAmulti{} and \snapshotBmulti{} include the multi-group implementation, whereas the other snapshots are post-processed from the gray RMHD snapshots. The notation ``n/a'' indicates that this model parameter was not included in the fit.}
\end{deluxetable*}\label{tab:ngc1313_fits}
%--------------------------------------------------------------------------//

\subsection{Example analysis: NGC 1313 X-1}
\label{s-results:ngc1313}
NGC 1313 X-1 is a well known ULX ($L_{\rm x} \sim 10^{40}\ \rm{erg} \ \rm{s}^{-1}$) located relatively nearby ($D \sim 4.2$ Mpc; \citealt{tullyetal2013}). The nature of its compact accretor is not currently known \citep{waltonetal2020}, although it has been suggested that changes observed in its hard X-ray flux might be consistent with a weakly magnetized neutron star \citep{middletonetal2023} entering a propeller state, where any pulsations (when not in propeller) are diluted due to scattering into the wind cone \citep{mushtukov2020}. Regardless of the nature of its compact object, an interesting feature of this source is the mostly stable shape of the hard X-ray spectrum $E \gtrsim 10$ keV as revealed by {\it NuSTAR} observations \citep{waltonetal2020}; this hard X-ray coverage makes NGC 1313 X-1 an excellent option for trialling our simulated spectral models.

Table \ref{tab:ngc1313_fits} shows the best-fit parameters for fits to the combined \textit{XMM-Newton} and \textit{NuSTAR} observations of NGC 1313 X-1 \citep{waltonetal2020}.  The data for NGC 1313 X-1 were provided by \cite{gurpideetal2021}.  For the \textit{NuSTAR} data we selected energies between $3-70$ keV, and for the \textit{XMM-Newton} data we used energies between $0.3-10$ keV.  In all of the fits, we allow two multiplicative constants to vary freely between the \textit{XMM-Newton} and \textit{NuSTAR} datasets to account for cross-calibration between the different detectors.  The FPMA detector constant was set to unity, while the two free constants yielded values $\lesssim1.38\pm0.2$. 

%% Reference fit (disbb+simpl(diskpbb))
We first highlight the ``Typical model'' column which includes two modified disk blackbody components and one hard X-ray component that is commonly used to fit this source \citep{Middleton2015b,Pinto2016,waltonetal2020,gurpideetal2021}.  To account for the hydrogen column along the line of sight for all model fits in this analysis, we include a neutral absorption component, \texttt{TBabs}, adopting the abundances from \cite{wilmsetal2000} and cross-sections from \cite{verneretal1996}. The column $N_{\rm H}$ was left free to vary (see \citealp{Middleton2015_winds,gurpideetal2021} for some discussion on the variability in the absorption column for this source).  The two modified disk blackbody components in this model combination are \texttt{diskbb} and \texttt{diskpbb}, used to model the softer and harder emission components, respectively.  The \texttt{diskpbb} component includes a free parameter, $p$, which describes the radial dependence of the local disk temperature, $T(r)\propto r^{-p}$.  When advection in the disk is considered important, such as in the case of super-Eddington accretion, the $p$ values are typically $p<0.75$ \citep{abramowiczetal1988}.  When $p=0.75$, the model recovers the thin disk \texttt{diskbb} solution.

Observations of NGC 1313 X-1 also show emission and absorption lines at energies $E \lesssim 2$ keV that are attributed to the presence of a mildly relativistic disk wind \citep{Middleton2014,Middleton2015_winds,Pinto2016,Pinto2020,gurpideetal2021}. Multiple Gaussian absorption components, \texttt{gabs}, are often included to account for some of these atomic features. We limit the \texttt{gabs} parameters to $E \leq 2 \; \rm keV$, line width $\sigma \leq 0.5 \; \rm keV$, and the line strength was allowed to be positive or negative to represent either emission or absorption.  In the Typical model, an additional component at moderate to high energies ($\gtrsim 10 \; \rm keV$) is included to capture the X-ray excess not adequately modeled by the multi-temperature disk components (see \citealt{waltonetal2020}).  We apply the same \texttt{simpl} convolution model to the \texttt{diskpbb} component as in previous works.  We set a lower limit on the power-law index parameter $\Gamma \geq 2$ as the uncertainties in the data at high energies $E \gtrsim 30 \; \rm keV$ cause \texttt{simpl} to return an unrealistically flat power-law.  The Typical model is written as: \texttt{TBabs}$\times$\texttt{gabs}$\times$(\texttt{diskbb}$+$(\texttt{simpl}$\times$\texttt{diskpbb})).  This model provides a reasonably good fit with $\chi^{2}=373.60$ for 341 degrees of freedom, comparable to the best-fits reported in \cite{Middleton2015,waltonetal2020,gurpideetal2021}.  One difference in our reproduction of this model is that we only included one \texttt{gabs} component with a line energy $E = 1 \; \rm keV$, line width of $\sigma = 0.01 \; \rm keV$, and line strength $N_{\rm{gabs}} = -0.02$.  In particular, this differs from the lines modeled in \cite{Middleton2015_winds} which were found at $E\simeq  0.66-0.74$ keV and $E \simeq 1.23-1.46$ keV.  If we restricted our model to include these specific lines, the fit returned $\chi^2 = 86.35$ for 338 degrees of freedom, which is a poorer $\chi^2$ than with a single \texttt{gabs} component. The overall fit is qualitatively similar between the two, however we noticed that the $f_{\rm sc}$ went to nearly $0$ if we used too many \texttt{gabs} components.   %$E = 1.0_{-0.03}^{+0.03}$ keV, line width of $\sigma=0.01_{-0.04}^{+0.05}$ keV, and line strength $N_{\rm{gabs}}=-0.02_{-0.02}^{+0.01}$. 

%% Spectral model fits
To compare to the Typical model, we replaced the \texttt{simpl$\times$diskpbb} component with one of the post-processed spectral models denoted in \texttt{XSPEC} as: \texttt{TBabs}$\times$\texttt{gabs}$\times$(\texttt{diskbb}$+$\texttt{MC spectrum}).  The \texttt{diskbb} component is included to model the soft X-ray flux absent from our spectral models. The spectral models have one fit parameter, $N = (10 \; {\rm kpc}/D)^2$, where $D$ is the distance to the source.  Assuming the distance to NGC 1313 X-1 is $D = 4.25 \; \rm Mpc$ \citep{tullyetal2013} gives a \texttt{MC spectrum} model normalization value of $N = 5.67 \times 10^{-6}$.  Table \ref{tab:ngc1313_fits} shows the best-fit \texttt{XSPEC} values for four snapshot models: \snapshotB{}, \snapshotAA{}, and their corresponding multi-group runs \snapshotBmulti{} and \snapshotAAmulti{}.  We also fit the other two spectral models from \snapshotC{} and \snapshotAB{}, but for the sake of brevity and lack of a multi-group counterpart for these snapshots, we do not include them in Tables \ref{tab:models} or \ref{tab:ngc1313_fits} but note that they provide poor fits to the data (the best-fit for \snapshotC{} returned a $\chi^2 = 637.87$ for $345$ degrees of freedom, and \snapshotAB{} returned a $\chi^{2} = 432.19$ for $345$ degrees of freedom).

%------------------------------  Figure  ----------------------------------
\begin{figure}[ht!]
    \centering
    \includegraphics[width=1.2\columnwidth]{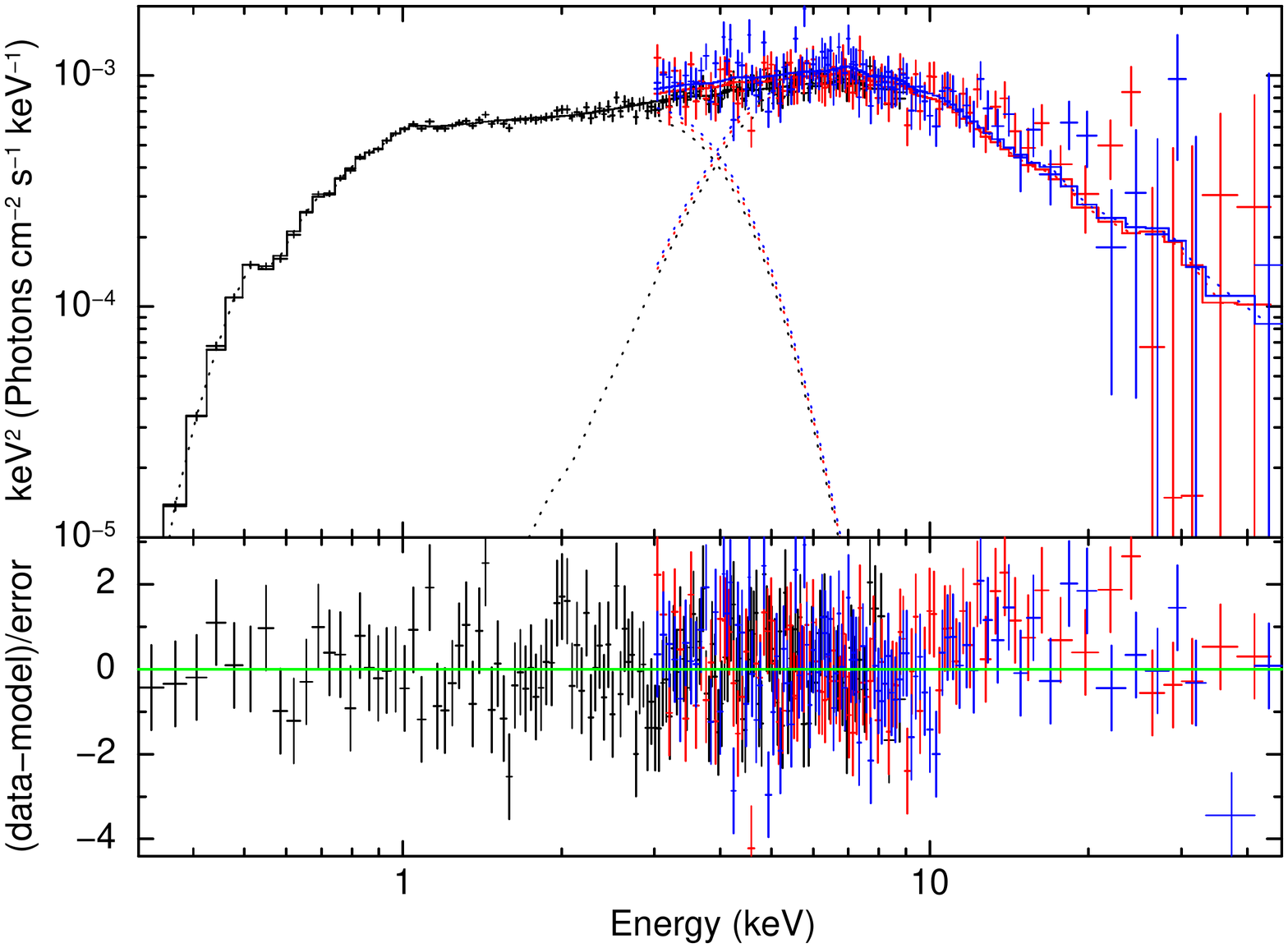}
    \caption{Best-fit to the combined \textit{XMM-Newton} (black data points) and \textit{NuSTAR} data (red and blue data points) of the ULX NGC 1313 X-1 using the post-processed spectral model from \snapshotAAmulti{}.  The top panel shows the spectral fit to the data with individual model components shown for \texttt{diskbb} (below 10 keV) and the \snapshotAAmulti{} model fitting the rest of the hard X-ray spectrum.  The bottom panel shows the fit residuals of the total model (green line) to the data.  The best-fit values are collected in Table \ref{tab:ngc1313_fits}.}
    \label{fig:fit-08250}
\end{figure}
%--------------------------------------------------------------------------//

All of the spectral model fits included a single \texttt{gabs} component except for \snapshotAAmulti{} which included two \texttt{gabs} components.  Most of the fits were insensitive to a second \texttt{gabs} component, but the $\chi^2$ for \snapshotAA{} improved from $\chi^2=433.20$ per 345 degrees of freedom with only one Gaussian absorption component to $\chi^2=394.50$ per 342 degrees of freedom with the addition of a second Gaussian component.  The two \texttt{gabs} components fit lines at $0.34 \; \rm keV$ with $\sigma = 0.49 \; \rm keV$, and a line at $0.67 \; \rm keV$ with $\sigma = 0.15 \; \rm keV$.  Generally, modeling the absorption and emission features of this source improves the $\chi^2$ of the fit residuals below $2$ keV, but it does not significantly impact the broader continuum fit. Thus we do not attempt to model these features in any detail as past studies have already done \citep{Middleton2015_winds,Pinto2016,Pinto2020,gurpideetal2021,Kosec2021}.

The $\Delta\chi^2$ improves significantly for fits with the multi-group models, as the harder X-ray tails in the models better match the observed \textit{NuSTAR} data.  
%------------------------------  Figure  ----------------------------------
\begin{figure}[ht]
    \centering
    \includegraphics[width=1.2\columnwidth]{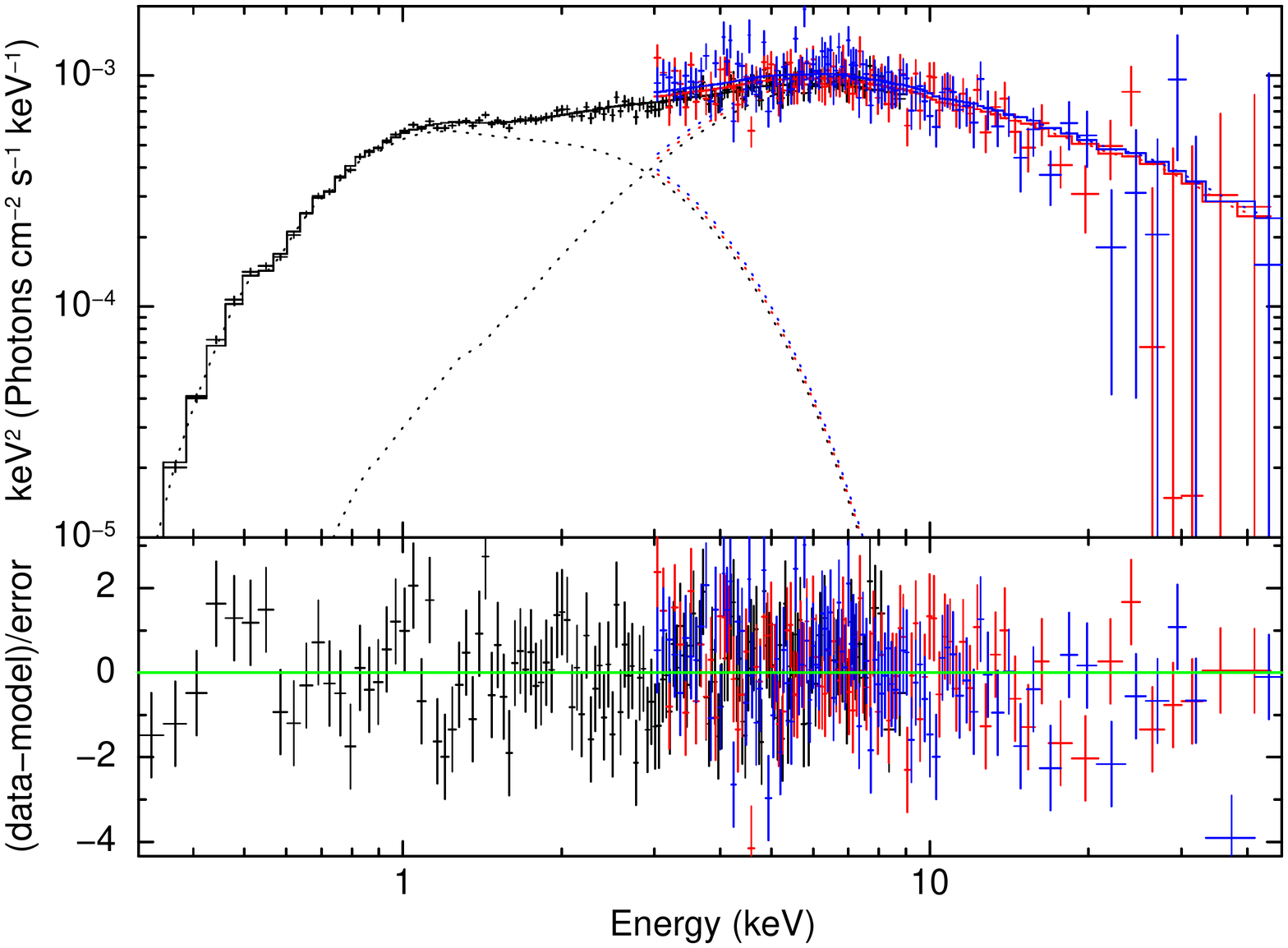}
    \caption{Same as Figure \ref{fig:fit-08250}, but using the \snapshotBmulti{} spectral model to fit the hard X-ray spectrum.}
    \label{fig:fit-18056}
\end{figure}
%--------------------------------------------------------------------------//
We show the two multi-group model combinations in Figures \ref{fig:fit-08250} and \ref{fig:fit-18056} for \snapshotAAmulti{} and \snapshotBmulti{}, respectively.  The individual model components are shown for the absorbed \texttt{diskbb} component below $10$ keV, and the component modeling the higher energy flux.  \snapshotAAmulti{} in Figure \ref{fig:fit-08250} is just slightly steeper than the\textit{NuSTAR} data at $\gtrsim 10$ keV, while \snapshotBmulti{} in Figure \ref{fig:fit-18056} is just slightly flatter at $\gtrsim 10$ keV.  Both spectral models, however, fit quite well in the 3-10 keV range.  Considering the simulated spectral component only has one parameter (the normalization), the deviation of the fit at $E \gtrsim 10$ keV qualitatively seems fairly reasonable.

%%%%%%%%%%%%%%%%%%%%%%%%%%%%%%%%%%%%%%%%%%%%%%%%%%%%%%%%%%%%%%%%%%%%%%%%%%%%%%%%%%%%%%%%%%%%%
%%%%%%%%%%%%%%%%%%%%%%%%%%%     DISCUSSION & CONCLUSIONS    %%%%%%%%%%%%%%%%%%%%%%%%%%%%%%%%% %%%%%%%%%%%%%%%%%%%%%%%%%%%%%%%%%%%%%%%%%%%%%%%%%%%%%%%%%%%%%%%%%%%%%%%%%%%%%%%%%%%%%%%%%%%%%
\section{Discussion}
\label{s-discussion}
Our results suggest that RMHD simulations can qualitatively reproduce the observed hard X-ray spectral shape seen in a number of ULX sources as long as the radiative heating/cooling associated with Compton scattering processes are well modeled. Nevertheless, the simulations presented here have only explored a limited range of parameter space and do not yet include all of the relevant physics.  Most importantly, the \Athena{} RMHD simulations neglect general relativistic effects such as light bending, relativistic beaming, and relativistic jets (although these simulations do generate radiatively driven disk winds).  New GRRMHD simulations are being performed with \Athena{} using the direct solutions of the radiation transfer equations  \citep{white2023} and we expect that the inclusion of general relativistic effects will have an impact on the accretion flow, disk structure, and associated spectral properties.  Post-processed spectra from such GRRMHD simulations will be the focus of future work.

In addition, we only include the inner 25\rg{} of these RMHD simulations in our MC spectral calculations as the simulation is not in steady state at larger radii. Thus, we do not accurately model the soft X-ray flux originating beyond $25$\rg{}.  Consequently, we only select photons coming out of a polar funnel angle, $\theta_{\rm f}$ to avoid the impact of photons which would normally interact with outer disk radii and become trapped in the disk or advected into the black hole.  Therefore, we stress that interpretations of these results be limited to the inner regions of the flow.

We also assume that protons and electrons are well coupled and so simulate a single temperature accretion flow $T=10^7$ K.  Some studies have suggested that two temperature accretion flows may become important in areas of low density, such as in the funnel regions \citep{liskaetal2022} and may result in softer X-ray spectra \citep{kinchetal2020}.  We compared the non-relativistic proton-electron relaxation timescale given by \cite{spitzer1956,stepney1983} assuming a single temperature for both the electrons and protons with the Compton timescale $t_{C} = (N_{\rm e} \sigma_{\rm T }c)^{-1}$ where $N_{\rm{e}}$ is the number density of electrons, $\sigma_{\rm{T}}$ is the Thomson cross-section, and $c$ is the speed of light.  We found that the relaxation time is much shorter than the Compton timescale and most other dynamical timescales for the temperature and densities in our simulation, except possibly in the very low density, high temperature region near the axis. Hence, our assumption of a single temperature for the protons and electrons seems self-consistent, but the single temperature assumption may need to be revisited in future work, particularly for simulations at lower accretion rates.

\subsection{Comparison with NGC 1313 X-1}
\label{s-discussion:models}
Fits to the \textit{XMM-Newton} and \textit{NuSTAR} data of NGC 1313 X-1 with the post-processed spectral models qualitatively reproduce the hard X-ray part of the spectrum, although the funnel luminosities for these spectra are at least an order of magnitude lower ($L_{\rm f}=1.3\times10^{39}\ \rm{erg}\ \rm{s}^{-1}$ for \snapshotAAmulti{}, and $L_{\rm f}=4.7\times10^{38}\ \rm{erg}\ \rm{s}^{-1}$ for \snapshotBmulti{}) than the observed luminosity ($L_{\rm{x}}\sim10^{40}\ \rm{erg}\ \rm{s}^{-1}$).  The implied distances from the spectral models are also much smaller ($D\sim1-2.13$ Mpc) when compared to the true distance to NGC 1313 X-1 of $D\simeq4.25$ Mpc \citep{tullyetal2013}.  This motivates future work with simulations at higher Eddington ratios and across a range of black hole masses, which might then better match the observed sources at their known distances.  Nevertheless, it is remarkable that a first principles calculation with only the normalization as a free parameter can provide best fitting $\chi^2$ values that are quantitatively competitive with commonly used phenomenological models.

\subsection{Comparison with previous work}
\label{s-discussion:previouswork}

Previous simulations have explored a range of accretion rates and masses, using a variety of setups both with and without general relativistic effects, finding radiative efficiencies that are both relatively large \citep[$\upeta \sim 5\%$, e.g.][]{jiangetal2014} or small \citep[$\upeta \lesssim 1\%$, e.g.][]{sadowskietal2014}.  The radiative efficiencies inferred directly from the gray RMHD simulations are typically a few percent, somewhat less than expected for thin accretion disks but not inconsistent with expectations for modestly super-Eddington accretion rates.  The radiative efficiencies $\upeta_{\rm f}= 1.13 - 2.56\%$ for the funnel region computed with MC post-processing are modestly lower than from the RMHD simulations for the gray snapshots.  In contrast, the snapshots produced from the multi-group calculations have slightly larger efficiencies ($\upeta_{\rm f} =1.92\%$ for \snapshotBmulti{}, and $\upeta_{\rm f}=3.34\%$ for \snapshotAAmulti{}), in better agreement with the luminosities directly inferred from RMHD simulations.

The Monte Carlo radiation transfer calculations performed in this work are similar to other post-processing codes, particularly those that use MC methods  \citep{dolence2009,schnittman2013,2021arXiv210805131K} that model Compton scattering and include general relativistic effects. The \texttt{HEROIC} code \citep{narayanetal2017}, which uses a combination of short and long characteristics instead of MC, provides similar capabilities. Although the \Athena{} module used here also supports general relativistic transfer, we treat the radiation transfer in Minkowski spacetime to be consistent with the non-relativistic simulations that generate the snapshots.  In contrast to \citet{kinchetal2019}, we do not currently perform any ionization calculations that would investigate atomic transitions.  We also do not use any integrated ray tracing algorithms that would integrate back along the photon path in our post-processing; although to create the images in Figure \ref{fig:images}, we extrapolate the photons escaping the MC domain out to a distant observer assuming flat spacetime.

Our approach compares most directly to those of \cite{narayanetal2017} and \citet{kitaki2017}, who consider the spectra produced from super-Eddington accretion simulations. Our results are broadly consistent with those of \citet{kitaki2017}, at least when one focuses on the hard component of the spectrum and face-on inclinations for the 10 $M_\odot$ black hole simulations. \cite{narayanetal2017} used \texttt{HEROIC} to post-process simulations from the GRRMHD code \texttt{KORAL} \citep{sadowskietal2013,sadowskietal2014,sadowski_narayan2015,sadowski_narayan2016}, which was used to simulate a broad range of super-Eddington accretion rates onto a $10 M_\odot$ black hole. They faced the same issue in that their GRRMHD simulations only reached inflow equilibrium out to a finite radius.  Instead of truncating the disk as we chose to do, they instead extrapolated the flow to larger radii using self-similar approximations.  This allowed them to explore the softer X-ray emission and angular dependence, but with the caveat that the outer regions of the calculation were not simulated directly.  We find that our spectra are more qualitatively consistent with their results when the gas temperatures in the \texttt{HERIOC} calculation were fixed to the values from \texttt{KORAL} (see green curves in Fig. 4 of \citealp{narayanetal2017}), but not consistent with their spectra after the radiation field and temperatures were self-consistently solved (red curves in the same figure).  The results from \texttt{HEROIC} show that their spectra become much softer after the temperature iteration, whilst our results suggest that a more self-consistent treatment of Compton cooling yields higher temperatures and harder spectra.  The origin of the difference is not clear to us, but we note that the \texttt{KORAL} simulations use a photon number conservation scheme that is different from what we use in our gray simulations.

\section{Summary}
\label{s-summary}
We present Monte Carlo post-processed spectral calculations of super-Eddington accretion onto a stellar-mass black hole from the \Athena{} RMHD simulation snapshots.  Our calculations suffer from two primary deficiencies.  We only achieve inflow equilibrium out to $\sim 25$ $r_{\rm g}$, which led us to truncate our spectral calculations at this radius.  Hence, the soft X-rays that come from the outer disk are absent.  If we instead include emission from the outer disk, it is significantly overestimated due to the cooling of the torus. Therefore we mainly focus on the hard X-ray spectrum in this work. These simulations also assume that the intensities follow a blackbody spectrum for the purposes of computing Compton cooling and mean opacities. Although this assumption is good for the optically thick disk, we find that using the blackbody assumption to estimate the average photon energy in the Compton cooling term is a poor approximation in the funnel regions where Comptonized electrons dominate the cooling.  This leads to an underestimate for the temperatures in the funnel for the gray RMHD simulations.  The underestimated temperatures produced spectra that were much softer and led to radiation energy densities above the disk being overestimated. We addressed this underestimate of the temperature by restarting the gray opacity simulations with a multi-group approach \citep{jiang2022} that treats Compton scattering with a Kompaneets-like source term. This produced simulation snapshots with higher temperatures in the spectral forming regions above the disk, leading to harder X-ray flux in better agreement with observed ULX spectra. 

We used phenomenological models to fit our Monte Carlo spectra. In most of the two-component (soft X-ray \& hard X-ray) models, the hard X-ray component was more accurately described with the \texttt{SIMPL} model compared to the power-law \texttt{POW} model, and yielded hard X-ray power-law slopes ranging from $\Gamma\sim2-4$ for spectra computed with gray RMHD snapshots.  The multi-group snapshot spectra tended to be fitted with flatter slopes, with$\Gamma\sim2-3$, comparable to the hard X-ray tails observed in NGC 1313 X-1 and Holmberg IX X-1 \citep{gurpideetal2021}.

Finally, we generated an \texttt{XSPEC} table model and directly fit our MC spectra to combined \textit{XMM-Newton} and \textit{NuSTAR} observations of the ULX NGC 1313 X-1. Despite only having one free parameter (the normalization), we find a good fit, which is competitive with the phenomenological models that are commonly used. Close inspection shows that the MC spectra provide a good fit at soft to moderately hard energies $E\lesssim10$ keV, but are either just slightly too steep in the case of \snapshotAAmulti{} or too flat in the case of \snapshotBmulti{} to exactly describe the hard X-ray power-law tail $E\gtrsim10$ keV.  The best-fit normalizations are also not consistent with the known distance to NGC 1313 X-1 and the model implies a lower luminosity than is observed. Although there are a number of caveats (such as the absence of general relativistic effects) and we have used only a single black hole mass and a relatively narrow range of accretion rates, this work nonetheless provides a promising direction for super-Eddington ULX accretion simulations, as these post-processed spectral models are close to describing the observed spectrum of NGC 1313 X-1. Simulations with the new GRRMHD implementation of the \Athena{} code \citep{white2023} are now exploring a range of masses and accretion rates, and post-processed spectra from these simulation will be presented in a future work.

\begin{acknowledgments}
This work was supported by NASA TCAN grant 80NSSC21K0496 and NASA ATP grant 80NSSC18K1018. BSM thanks the Jefferson Scholars Foundation Graduate Fellowship in support of this work. Part of this work was performed using resources provided by the Cambridge Service for Data Driven Discovery (CSD3) operated by the University of Cambridge Research Computing Service (www.csd3.cam.ac.uk), provided by Dell EMC and Intel using Tier-2 funding from the Engineering and Physical Sciences Research Council (capital grant EP/T022159/1), and DiRAC funding from the Science and Technology Facilities Council (www.dirac.ac.uk). Resources supporting this work were also provided by the High-End Computing (HEC) program through the NASA Advanced Supercomputing (NAS) Division at Ames Research Center. The Center for Computational Astrophysics at the Flatiron Institute is supported by the Simons Foundation.
\end{acknowledgments}

%%%%%%%%%%%%%%%%%%%%%%%%%%%%%%%%%%%%%%%%%%%%%%%%%%%%%%%%%%%%%%%%%%%%%%%%%%%%%%%%%%%%%%%%%%%%%
%%%%%%%%%%%%%%%%%%%%%%%%%%     ACKNOWLEDGEMENTS & BIBLIOGRAPHY    %%%%%%%%%%%%%%%%%%%%%%%%%%% %%%%%%%%%%%%%%%%%%%%%%%%%%%%%%%%%%%%%%%%%%%%%%%%%%%%%%%%%%%%%%%%%%%%%%%%%%%%%%%%%%%%%%%%%%%%%
\bibliography{bibliography.bib}

\begin{thebibliography}{}
\expandafter\ifx\csname natexlab\endcsname\relax\def\natexlab#1{#1}\fi
\providecommand{\url}[1]{\href{#1}{#1}}
\providecommand{\dodoi}[1]{doi:~\href{http://doi.org/#1}{\nolinkurl{#1}}}
\providecommand{\doeprint}[1]{\href{http://ascl.net/#1}{\nolinkurl{http://ascl.net/#1}}}
\providecommand{\doarXiv}[1]{\href{https://arxiv.org/abs/#1}{\nolinkurl{https://arxiv.org/abs/#1}}}

\bibitem[{{Abramowicz} {et~al.}(1988){Abramowicz}, {Czerny}, {Lasota}, \&
  {Szuszkiewicz}}]{abramowiczetal1988}
{Abramowicz}, M.~A., {Czerny}, B., {Lasota}, J.~P., \& {Szuszkiewicz}, E. 1988,
  \apj, 332, 646, \dodoi{10.1086/166683}

\bibitem[{{Arnaud}(1996)}]{arnaud1996}
{Arnaud}, K.~A. 1996, in Astronomical Society of the Pacific Conference Series,
  Vol. 101, Astronomical Data Analysis Software and Systems V, ed. G.~H.
  {Jacoby} \& J.~{Barnes}, 17

\bibitem[{Asahina {et~al.}(2020)Asahina, Takahashi, \&
  Ohsuga}]{asahinaetal2020}
Asahina, Y., Takahashi, H.~R., \& Ohsuga, K. 2020, The Astrophysical Journal,
  901, 96, \dodoi{10.3847/1538-4357/abaf51}

\bibitem[{{Bachetti} {et~al.}(2014){Bachetti}, {Harrison}, {Walton},
  {Grefenstette}, {Chakrabarty}, {F{\"u}rst}, {Barret}, {Beloborodov}, {Boggs},
  {Christensen}, {Craig}, {Fabian}, {Hailey}, {Hornschemeier}, {Kaspi},
  {Kulkarni}, {Maccarone}, {Miller}, {Rana}, {Stern}, {Tendulkar}, {Tomsick},
  {Webb}, \& {Zhang}}]{bachettietal2014}
{Bachetti}, M., {Harrison}, F.~A., {Walton}, D.~J., {et~al.} 2014, \nat, 514,
  202, \dodoi{10.1038/nature13791}

\bibitem[{{Balbus} \& {Hawley}(1991)}]{balbushawley1991}
{Balbus}, S.~A., \& {Hawley}, J.~F. 1991, \apj, 376, 214,
  \dodoi{10.1086/170270}

\bibitem[{{Brightman} {et~al.}(2016){Brightman}, {Harrison}, {Barret}, {Davis},
  {F{\"u}rst}, {Madsen}, {Middleton}, {Miller}, {Stern}, {Tao}, \&
  {Walton}}]{brightmanetal2016}
{Brightman}, M., {Harrison}, F.~A., {Barret}, D., {et~al.} 2016, \apj, 829, 28,
  \dodoi{10.3847/0004-637X/829/1/28}

\bibitem[{{Chandrasekhar}(1960)}]{chandrasekhar1960}
{Chandrasekhar}, S. 1960, {Radiative transfer}

\bibitem[{Dage {et~al.}(2021)Dage, Vowell, Thygesen, Bahramian, Haggard,
  Kovlakas, Kundu, Maccarone, Strader, Urquhart, \& Zepf}]{dage2021}
Dage, K.~C., Vowell, N., Thygesen, E., {et~al.} 2021, Monthly Notices of the
  Royal Astronomical Society, 508, 4008, \dodoi{10.1093/mnras/stab2850}

\bibitem[{{Davis} {et~al.}(2009){Davis}, {Blaes}, {Hirose}, \&
  {Krolik}}]{davisetal2009}
{Davis}, S.~W., {Blaes}, O.~M., {Hirose}, S., \& {Krolik}, J.~H. 2009, \apj,
  703, 569, \dodoi{10.1088/0004-637X/703/1/569}

\bibitem[{{Davis} {et~al.}(2012){Davis}, {Stone}, \& {Jiang}}]{davisetal2012}
{Davis}, S.~W., {Stone}, J.~M., \& {Jiang}, Y.-F. 2012, \apjs, 199, 9,
  \dodoi{10.1088/0067-0049/199/1/9}

\bibitem[{{Dolence} {et~al.}(2009){Dolence}, {Gammie}, {Mo{\'s}cibrodzka}, \&
  {Leung}}]{dolence2009}
{Dolence}, J.~C., {Gammie}, C.~F., {Mo{\'s}cibrodzka}, M., \& {Leung}, P.~K.
  2009, \apjs, 184, 387, \dodoi{10.1088/0067-0049/184/2/387}

\bibitem[{{Earnshaw}(2016)}]{earnshaw2016}
{Earnshaw}, H.~M. 2016, Astronomische Nachrichten, 337, 448,
  \dodoi{10.1002/asna.201612328}

\bibitem[{{Event Horizon Telescope Collaboration} {et~al.}(2019){Event Horizon
  Telescope Collaboration}, {Akiyama}, {Alberdi}, {Alef}, {Asada}, {Azulay},
  {Baczko}, {Ball}, {Balokovi{\'c}}, {Barrett}, {Bintley}, {Blackburn},
  {Boland}, {Bouman}, {Bower}, {Bremer}, {Brinkerink}, {Brissenden}, {Britzen},
  {Broderick}, {Broguiere}, {Bronzwaer}, {Byun}, {Carlstrom}, {Chael}, {Chan},
  {Chatterjee}, {Chatterjee}, {Chen}, {Chen}, {Cho}, {Christian}, {Conway},
  {Cordes}, {Crew}, {Cui}, {Davelaar}, {De Laurentis}, {Deane}, {Dempsey},
  {Desvignes}, {Dexter}, {Doeleman}, {Eatough}, {Falcke}, {Fish}, {Fomalont},
  {Fraga-Encinas}, {Friberg}, {Fromm}, {G{\'o}mez}, {Galison}, {Gammie},
  {Garc{\'\i}a}, {Gentaz}, {Georgiev}, {Goddi}, {Gold}, {Gu}, {Gurwell},
  {Hada}, {Hecht}, {Hesper}, {Ho}, {Ho}, {Honma}, {Huang}, {Huang}, {Hughes},
  {Ikeda}, {Inoue}, {Issaoun}, {James}, {Jannuzi}, {Janssen}, {Jeter}, {Jiang},
  {Johnson}, {Jorstad}, {Jung}, {Karami}, {Karuppusamy}, {Kawashima},
  {Keating}, {Kettenis}, {Kim}, {Kim}, {Kim}, {Kino}, {Koay}, {Koch}, {Koyama},
  {Kramer}, {Kramer}, {Krichbaum}, {Kuo}, {Lauer}, {Lee}, {Li}, {Li},
  {Lindqvist}, {Liu}, {Liuzzo}, {Lo}, {Lobanov}, {Loinard}, {Lonsdale}, {Lu},
  {MacDonald}, {Mao}, {Markoff}, {Marrone}, {Marscher}, {Mart{\'\i}-Vidal},
  {Matsushita}, {Matthews}, {Medeiros}, {Menten}, {Mizuno}, {Mizuno}, {Moran},
  {Moriyama}, {Moscibrodzka}, {Mul{\ensuremath{\ddot{}}}ler}, {Nagai}, {Nagar},
  {Nakamura}, {Narayan}, {Narayanan}, {Natarajan}, {Neri}, {Ni}, {Noutsos},
  {Okino}, {Olivares}, {Oyama}, {{\"O}zel}, {Palumbo}, {Patel}, {Pen}, {Pesce},
  {Pi{\'e}tu}, {Plambeck}, {PopStefanija}, {Porth}, {Prather},
  {Preciado-L{\'o}pez}, {Psaltis}, {Pu}, {Ramakrishnan}, {Rao}, {Rawlings},
  {Raymond}, {Rezzolla}, {Ripperda}, {Roelofs}, {Rogers}, {Ros}, {Rose},
  {Roshanineshat}, {Rottmann}, {Roy}, {Ruszczyk}, {Ryan}, {Rygl},
  {S{\'a}nchez}, {S{\'a}nchez-Arguelles}, {Sasada}, {Savolainen}, {Schloerb},
  {Schuster}, {Shao}, {Shen}, {Small}, {Sohn}, {SooHoo}, {Tazaki}, {Tiede},
  {Tilanus}, {Titus}, {Toma}, {Torne}, {Trent}, {Trippe}, {Tsuda}, {van
  Bemmel}, {van Langevelde}, {van Rossum}, {Wagner}, {Wardle}, {Weintroub},
  {Wex}, {Wharton}, {Wielgus}, {Wong}, {Wu}, {Young}, {Young}, {Younsi},
  {Yuan}, {Yuan}, {Zensus}, {Zhao}, {Zhao}, {Zhu}, {Anczarski}, {Baganoff},
  {Eckart}, {Farah}, {Haggard}, {Meyer-Zhao}, {Michalik}, {Nadolski},
  {Neilsen}, {Nishioka}, {Nowak}, {Pradel}, {Primiani}, {Souccar},
  {Vertatschitsch}, {Yamaguchi}, \& {Zhang}}]{eht2019}
{Event Horizon Telescope Collaboration}, {Akiyama}, K., {Alberdi}, A., {et~al.}
  2019, \apjl, 875, L5, \dodoi{10.3847/2041-8213/ab0f43}

\bibitem[{{Event Horizon Telescope Collaboration} {et~al.}(2022){Event Horizon
  Telescope Collaboration}, {Akiyama}, {Alberdi}, {Alef}, {Algaba}, {Anantua},
  {Asada}, {Azulay}, {Bach}, {Baczko}, {Ball}, {Balokovi{\'c}}, {Barrett},
  {Baub{\"o}ck}, {Benson}, {Bintley}, {Blackburn}, {Blundell}, {Bouman},
  {Bower}, {Boyce}, {Bremer}, {Brinkerink}, {Brissenden}, {Britzen},
  {Broderick}, {Broguiere}, {Bronzwaer}, {Bustamante}, {Byun}, {Carlstrom},
  {Ceccobello}, {Chael}, {Chan}, {Chatterjee}, {Chatterjee}, {Chen}, {Chen},
  {Cheng}, {Cho}, {Christian}, {Conroy}, {Conway}, {Cordes}, {Crawford},
  {Crew}, {Cruz-Osorio}, {Cui}, {Davelaar}, {De Laurentis}, {Deane}, {Dempsey},
  {Desvignes}, {Dexter}, {Dhruv}, {Doeleman}, {Dougal}, {Dzib}, {Eatough},
  {Emami}, {Falcke}, {Farah}, {Fish}, {Fomalont}, {Ford}, {Fraga-Encinas},
  {Freeman}, {Friberg}, {Fromm}, {Fuentes}, {Galison}, {Gammie}, {Garc{\'\i}a},
  {Gentaz}, {Georgiev}, {Goddi}, {Gold}, {G{\'o}mez-Ruiz}, {G{\'o}mez}, {Gu},
  {Gurwell}, {Hada}, {Haggard}, {Haworth}, {Hecht}, {Hesper}, {Heumann}, {Ho},
  {Ho}, {Honma}, {Huang}, {Huang}, {Hughes}, {Ikeda}, {Violette Impellizzeri},
  {Inoue}, {Issaoun}, {James}, {Jannuzi}, {Janssen}, {Jeter}, {Jiang},
  {Jim{\'e}nez-Rosales}, {Johnson}, {Jorstad}, {Joshi}, {Jung}, {Karami},
  {Karuppusamy}, {Kawashima}, {Keating}, {Kettenis}, {Kim}, {Kim}, {Kim},
  {Kim}, {Kino}, {Koay}, {Kocherlakota}, {Kofuji}, {Koch}, {Koyama}, {Kramer},
  {Kramer}, {Krichbaum}, {Kuo}, {La Bella}, {Lauer}, {Lee}, {Lee}, {Leung},
  {Levis}, {Li}, {Lico}, {Lindahl}, {Lindqvist}, {Lisakov}, {Liu}, {Liu},
  {Liuzzo}, {Lo}, {Lobanov}, {Loinard}, {Lonsdale}, {Lu}, {Mao}, {Marchili},
  {Markoff}, {Marrone}, {Marscher}, {Mart{\'\i}-Vidal}, {Matsushita},
  {Matthews}, {Medeiros}, {Menten}, {Michalik}, {Mizuno}, {Mizuno}, {Moran},
  {Moriyama}, {Moscibrodzka}, {M{\"u}ller}, {Mus}, {Musoke}, {Myserlis},
  {Nadolski}, {Nagai}, {Nagar}, {Nakamura}, {Narayan}, {Narayanan},
  {Natarajan}, {Nathanail}, {Navarro Fuentes}, {Neilsen}, {Neri}, {Ni},
  {Noutsos}, {Nowak}, {Oh}, {Okino}, {Olivares}, {Ortiz-Le{\'o}n}, {Oyama},
  {{\"O}zel}, {Palumbo}, {Filippos Paraschos}, {Park}, {Parsons}, {Patel},
  {Pen}, {Pesce}, {Pi{\'e}tu}, {Plambeck}, {PopStefanija}, {Porth},
  {P{\"o}tzl}, {Prather}, {Preciado-L{\'o}pez}, {Psaltis}, {Pu},
  {Ramakrishnan}, {Rao}, {Rawlings}, {Raymond}, {Rezzolla}, {Ricarte},
  {Ripperda}, {Roelofs}, {Rogers}, {Ros}, {Romero-Ca{\~n}izales},
  {Roshanineshat}, {Rottmann}, {Roy}, {Ruiz}, {Ruszczyk}, {Rygl},
  {S{\'a}nchez}, {S{\'a}nchez-Arg{\"u}elles}, {S{\'a}nchez-Portal}, {Sasada},
  {Satapathy}, {Savolainen}, {Schloerb}, {Schonfeld}, {Schuster}, {Shao},
  {Shen}, {Small}, {Sohn}, {SooHoo}, {Souccar}, {Sun}, {Tazaki}, {Tetarenko},
  {Tiede}, {Tilanus}, {Titus}, {Torne}, {Traianou}, {Trent}, {Trippe}, {Turk},
  {van Bemmel}, {van Langevelde}, {van Rossum}, {Vos}, {Wagner},
  {Ward-Thompson}, {Wardle}, {Weintroub}, {Wex}, {Wharton}, {Wielgus}, {Wiik},
  {Witzel}, {Wondrak}, {Wong}, {Wu}, {Yamaguchi}, {Yoon}, {Young}, {Young},
  {Younsi}, {Yuan}, {Yuan}, {Zensus}, {Zhang}, {Zhao}, {Zhao}, {Chan}, {Qiu},
  {Ressler}, \& {White}}]{eht2022}
---. 2022, \apjl, 930, L16, \dodoi{10.3847/2041-8213/ac6672}

\bibitem[{{Farrell} {et~al.}(2009){Farrell}, {Webb}, {Barret}, {Godet}, \&
  {Rodrigues}}]{farrelletal2009}
{Farrell}, S.~A., {Webb}, N.~A., {Barret}, D., {Godet}, O., \& {Rodrigues},
  J.~M. 2009, \nat, 460, 73, \dodoi{10.1038/nature08083}

\bibitem[{Fragile {et~al.}(2014)Fragile, Olejar, \& Anninos}]{fragileetal2014}
Fragile, P.~C., Olejar, A., \& Anninos, P. 2014, The Astrophysical Journal,
  796, 22, \dodoi{10.1088/0004-637X/796/1/22}

\bibitem[{Gladstone {et~al.}(2009)Gladstone, Roberts, \&
  Done}]{gladstoneetal2009}
Gladstone, J.~C., Roberts, T.~P., \& Done, C. 2009, Monthly Notices of the
  Royal Astronomical Society, 397, 1836,
  \dodoi{10.1111/j.1365-2966.2009.15123.x}

\bibitem[{{Gonz\'alez, M.} {et~al.}(2007){Gonz\'alez, M.}, {Audit, E.}, \&
  {Huynh, P.}}]{gonzalezetal2007}
{Gonz\'alez, M.}, {Audit, E.}, \& {Huynh, P.} 2007, A\&A, 464, 429,
  \dodoi{10.1051/0004-6361:20065486}

\bibitem[{{G{\'u}rpide} {et~al.}(2021){G{\'u}rpide}, {Godet}, {Koliopanos},
  {Webb}, \& {Olive}}]{gurpideetal2021}
{G{\'u}rpide}, A., {Godet}, O., {Koliopanos}, F., {Webb}, N., \& {Olive}, J.~F.
  2021, \aap, 649, A104, \dodoi{10.1051/0004-6361/202039572}

\bibitem[{Howell \& Greenough(2003)}]{howellgreenough2003}
Howell, L.~H., \& Greenough, J.~A. 2003, Journal of Computational Physics, 184,
  53, \dodoi{https://doi.org/10.1016/S0021-9991(02)00015-3}

\bibitem[{{Huang} {et~al.}(2023){Huang}, {Jiang}, {Feng}, {Davis}, {Stone}, \&
  {Middleton}}]{huangetal2023}
{Huang}, J., {Jiang}, Y.-F., {Feng}, H., {et~al.} 2023, arXiv e-prints,
  arXiv:2301.12679, \dodoi{10.48550/arXiv.2301.12679}

\bibitem[{{Jiang}(2021)}]{jiang2021}
{Jiang}, Y.-F. 2021, \apjs, 253, 49, \dodoi{10.3847/1538-4365/abe303}

\bibitem[{Jiang(2022)}]{jiang2022}
Jiang, Y.-F. 2022, The Astrophysical Journal Supplement Series, 263, 4,
  \dodoi{10.3847/1538-4365/ac9231}

\bibitem[{{Jiang} {et~al.}(2012){Jiang}, {Stone}, \& {Davis}}]{jiangetal2012}
{Jiang}, Y.~F., {Stone}, J.~M., \& {Davis}, S.~W. 2012, in Astronomical Society
  of the Pacific Conference Series, Vol. 459, Numerical Modeling of Space
  Plasma Slows (ASTRONUM 2011), ed. N.~V. {Pogorelov}, J.~A. {Font},
  E.~{Audit}, \& G.~P. {Zank}, 322

\bibitem[{{Jiang} {et~al.}(2014){Jiang}, {Stone}, \& {Davis}}]{jiangetal2014}
{Jiang}, Y.-F., {Stone}, J.~M., \& {Davis}, S.~W. 2014, \apj, 796, 106,
  \dodoi{10.1088/0004-637X/796/2/106}

\bibitem[{{Jiang} {et~al.}(2019){Jiang}, {Stone}, \& {Davis}}]{jiang2019super}
---. 2019, \apj, 880, 67, \dodoi{10.3847/1538-4357/ab29ff}

\bibitem[{{Kawashima} {et~al.}(2012){Kawashima}, {Ohsuga}, {Mineshige},
  {Yoshida}, {Heinzeller}, \& {Matsumoto}}]{kawashima2012}
{Kawashima}, T., {Ohsuga}, K., {Mineshige}, S., {et~al.} 2012, \apj, 752, 18,
  \dodoi{10.1088/0004-637X/752/1/18}

\bibitem[{{Kawashima} {et~al.}(2021){Kawashima}, {Ohsuga}, \&
  {Takahashi}}]{2021arXiv210805131K}
{Kawashima}, T., {Ohsuga}, K., \& {Takahashi}, H.~R. 2021, arXiv e-prints,
  arXiv:2108.05131, \dodoi{10.48550/arXiv.2108.05131}

\bibitem[{{Kinch} {et~al.}(2020){Kinch}, {Noble}, {Schnittman}, \&
  {Krolik}}]{kinchetal2020}
{Kinch}, B.~E., {Noble}, S.~C., {Schnittman}, J.~D., \& {Krolik}, J.~H. 2020,
  \apj, 904, 117, \dodoi{10.3847/1538-4357/abc176}

\bibitem[{{Kinch} {et~al.}(2019){Kinch}, {Schnittman}, {Kallman}, \&
  {Krolik}}]{kinchetal2019}
{Kinch}, B.~E., {Schnittman}, J.~D., {Kallman}, T.~R., \& {Krolik}, J.~H. 2019,
  \apj, 873, 71, \dodoi{10.3847/1538-4357/ab05d5}

\bibitem[{{Kinch} {et~al.}(2021){Kinch}, {Schnittman}, {Noble}, {Kallman}, \&
  {Krolik}}]{kinchetal2021}
{Kinch}, B.~E., {Schnittman}, J.~D., {Noble}, S.~C., {Kallman}, T.~R., \&
  {Krolik}, J.~H. 2021, \apj, 922, 270, \dodoi{10.3847/1538-4357/ac2b9a}

\bibitem[{{King} {et~al.}(2023){King}, {Lasota}, \& {Middleton}}]{KLM2023}
{King}, A., {Lasota}, J.-P., \& {Middleton}, M. 2023, \nar, 96, 101672,
  \dodoi{10.1016/j.newar.2022.101672}

\bibitem[{{Kitaki} {et~al.}(2017){Kitaki}, {Mineshige}, {Ohsuga}, \&
  {Kawashima}}]{kitaki2017}
{Kitaki}, T., {Mineshige}, S., {Ohsuga}, K., \& {Kawashima}, T. 2017, \pasj,
  69, 92, \dodoi{10.1093/pasj/psx101}

\bibitem[{{Kosec} {et~al.}(2021){Kosec}, {Pinto}, {Reynolds}, {Guainazzi},
  {Kara}, {Walton}, {Fabian}, {Parker}, \& {Valtchanov}}]{Kosec2021}
{Kosec}, P., {Pinto}, C., {Reynolds}, C.~S., {et~al.} 2021, \mnras, 508, 3569,
  \dodoi{10.1093/mnras/stab2856}

\bibitem[{Krumholz {et~al.}(2007)Krumholz, Klein, McKee, \&
  Bolstad}]{krumholzetal2007}
Krumholz, M.~R., Klein, R.~I., McKee, C.~F., \& Bolstad, J. 2007, The
  Astrophysical Journal, 667, 626, \dodoi{10.1086/520791}

\bibitem[{{Kubota} \& {Done}(2019)}]{kubota2019}
{Kubota}, A., \& {Done}, C. 2019, \mnras, 489, 524,
  \dodoi{10.1093/mnras/stz2140}

\bibitem[{{Levermore}(1984)}]{levermore1984}
{Levermore}, C.~D. 1984, \jqsrt, 31, 149, \dodoi{10.1016/0022-4073(84)90112-2}

\bibitem[{{Liska} {et~al.}(2022){Liska}, {Musoke}, {Tchekhovskoy}, {Porth}, \&
  {Beloborodov}}]{liskaetal2022}
{Liska}, M.~T.~P., {Musoke}, G., {Tchekhovskoy}, A., {Porth}, O., \&
  {Beloborodov}, A.~M. 2022, \apjl, 935, L1, \dodoi{10.3847/2041-8213/ac84db}

\bibitem[{{McKinney} {et~al.}(2014){McKinney}, {Tchekhovskoy}, \&
  {Narayan}}]{mckinneyetal2014}
{McKinney}, J.~C., {Tchekhovskoy}, {S{\k{a}}dowski}, A., \& {Narayan}, R. 2014,
  \mnras, 441, 3177, \dodoi{10.1093/mnras/stu762}

\bibitem[{{Menon} {et~al.}(2022){Menon}, {Federrath}, {Krumholz}, {Kuiper},
  {Wibking}, \& {Jung}}]{menonetal2022}
{Menon}, S.~H., {Federrath}, C., {Krumholz}, M.~R., {et~al.} 2022, \mnras, 512,
  401, \dodoi{10.1093/mnras/stac485}

\bibitem[{{Mezcua} {et~al.}(2013){Mezcua}, {Roberts}, {Sutton}, \&
  {Lobanov}}]{mezcuaetal2013}
{Mezcua}, M., {Roberts}, T.~P., {Sutton}, A.~D., \& {Lobanov}, A.~P. 2013,
  \mnras, 436, 3128, \dodoi{10.1093/mnras/stt1794}

\bibitem[{{Middleton} {et~al.}(2023){Middleton}, {G{\'u}rpide}, \&
  {Walton}}]{middletonetal2023}
{Middleton}, M., {G{\'u}rpide}, A., \& {Walton}, D.~J. 2023, \mnras, 519, 2224,
  \dodoi{10.1093/mnras/stac3380}

\bibitem[{{Middleton} {et~al.}(2015{\natexlab{a}}){Middleton}, {Heil},
  {Pintore}, {Walton}, \& {Roberts}}]{Middleton2015}
{Middleton}, M.~J., {Heil}, L., {Pintore}, F., {Walton}, D.~J., \& {Roberts},
  T.~P. 2015{\natexlab{a}}, \mnras, 447, 3243, \dodoi{10.1093/mnras/stu2644}

\bibitem[{{Middleton} {et~al.}(2015{\natexlab{b}}){Middleton}, {Walton},
  {Fabian}, {Roberts}, {Heil}, {Pinto}, {Anderson}, \&
  {Sutton}}]{Middleton2015b}
{Middleton}, M.~J., {Walton}, D.~J., {Fabian}, A., {et~al.} 2015{\natexlab{b}},
  \mnras, 454, 3134, \dodoi{10.1093/mnras/stv2214}

\bibitem[{{Middleton} {et~al.}(2015{\natexlab{c}}){Middleton}, {Walton},
  {Fabian}, {Roberts}, {Heil}, {Pinto}, {Anderson}, \&
  {Sutton}}]{Middleton2015_winds}
---. 2015{\natexlab{c}}, \mnras, 454, 3134, \dodoi{10.1093/mnras/stv2214}

\bibitem[{{Middleton} {et~al.}(2014){Middleton}, {Walton}, {Roberts}, \&
  {Heil}}]{Middleton2014}
{Middleton}, M.~J., {Walton}, D.~J., {Roberts}, T.~P., \& {Heil}, L. 2014,
  \mnras, 438, L51, \dodoi{10.1093/mnrasl/slt157}

\bibitem[{Miller {et~al.}(2004)Miller, Fabian, \& Miller}]{milleretal2004}
Miller, J.~M., Fabian, A.~C., \& Miller, M.~C. 2004, The Astrophysical Journal,
  614, L117, \dodoi{10.1086/425316}

\bibitem[{{Mitsuda} {et~al.}(1984){Mitsuda}, {Inoue}, {Koyama}, {Makishima},
  {Matsuoka}, {Ogawara}, {Shibazaki}, {Suzuki}, {Tanaka}, \&
  {Hirano}}]{mitsudaetal1984}
{Mitsuda}, K., {Inoue}, H., {Koyama}, K., {et~al.} 1984, \pasj, 36, 741

\bibitem[{{Moens, N.} {et~al.}(2022){Moens, N.}, {Sundqvist, J. O.}, {El
  Mellah, I.}, {Poniatowski, L.}, {Teunissen, J.}, \& {Keppens,
  R.}}]{moensetal2022}
{Moens, N.}, {Sundqvist, J. O.}, {El Mellah, I.}, {et~al.} 2022, A\&A, 657,
  A81, \dodoi{10.1051/0004-6361/202141023}

\bibitem[{Mushtukov {et~al.}(2020)Mushtukov, Portegies Zwart, Tsygankov,
  Nagirner, \& Poutanen}]{mushtukov2020}
Mushtukov, A.~A., Portegies Zwart, S., Tsygankov, S.~S., Nagirner, D.~I., \&
  Poutanen, J. 2020, Monthly Notices of the Royal Astronomical Society, 501,
  2424, \dodoi{10.1093/mnras/staa3809}

\bibitem[{Narayan {et~al.}(2017)Narayan, {S{\k{a}}dowski}, \&
  Soria}]{narayanetal2017}
Narayan, R., {S{\k{a}}dowski}, A., \& Soria, R. 2017, Monthly Notices of the
  Royal Astronomical Society, 469, 2997, \dodoi{10.1093/mnras/stx1027}

\bibitem[{{Ohsuga} \& {Mineshige}(2011)}]{ohsuga_mineshige2011}
{Ohsuga}, K., \& {Mineshige}, S. 2011, \apj, 736, 2,
  \dodoi{10.1088/0004-637X/736/1/2}

\bibitem[{Ohsuga {et~al.}(2005)Ohsuga, Mori, Nakamoto, \&
  Mineshige}]{ohsugaetal2005}
Ohsuga, K., Mori, M., Nakamoto, T., \& Mineshige, S. 2005, The Astrophysical
  Journal, 628, 368, \dodoi{10.1086/430728}

\bibitem[{{Oskinova} {et~al.}(2019){Oskinova}, {Bik}, {Mas-Hesse}, {Hayes},
  {Adamo}, {{\"O}stlin}, {F{\"u}rst}, \&
  {Ot{\'\i}-Floranes}}]{oskinovaetal2019}
{Oskinova}, L.~M., {Bik}, A., {Mas-Hesse}, J.~M., {et~al.} 2019, \aap, 627,
  A63, \dodoi{10.1051/0004-6361/201935414}

\bibitem[{{Paczy{\'n}sky} \& {Wiita}(1980)}]{paczynskywiita1980}
{Paczy{\'n}sky}, B., \& {Wiita}, P.~J. 1980, \aap, 88, 23

\bibitem[{{Pinto} {et~al.}(2016){Pinto}, {Middleton}, \& {Fabian}}]{Pinto2016}
{Pinto}, C., {Middleton}, M.~J., \& {Fabian}, A.~C. 2016, \nat, 533, 64,
  \dodoi{10.1038/nature17417}

\bibitem[{{Pinto} \& {Walton}(2023)}]{pinto_walton2023}
{Pinto}, C., \& {Walton}, D.~J. 2023, arXiv e-prints, arXiv:2302.00006,
  \dodoi{10.48550/arXiv.2302.00006}

\bibitem[{{Pinto} {et~al.}(2020){Pinto}, {Walton}, {Kara}, {Parker}, {Soria},
  {Kosec}, {Middleton}, {Alston}, {Fabian}, {Guainazzi}, {Roberts}, {Fuerst},
  {Earnshaw}, {Sathyaprakash}, \& {Barret}}]{Pinto2020}
{Pinto}, C., {Walton}, D.~J., {Kara}, E., {et~al.} 2020, \mnras, 492, 4646,
  \dodoi{10.1093/mnras/staa118}

\bibitem[{Pintore {et~al.}(2017)Pintore, Zampieri, Stella, Wolter, Mereghetti,
  \& Israel}]{pintoreetal2017}
Pintore, F., Zampieri, L., Stella, L., {et~al.} 2017, The Astrophysical
  Journal, 836, 113, \dodoi{10.3847/1538-4357/836/1/113}

\bibitem[{{Poutanen} {et~al.}(2007){Poutanen}, {Lipunova}, {Fabrika},
  {Butkevich}, \& {Abolmasov}}]{Poutanen2007}
{Poutanen}, J., {Lipunova}, G., {Fabrika}, S., {Butkevich}, A.~G., \&
  {Abolmasov}, P. 2007, \mnras, 377, 1187,
  \dodoi{10.1111/j.1365-2966.2007.11668.x}

\bibitem[{{Pozdnyakov} {et~al.}(1983){Pozdnyakov}, {Sobol}, \&
  {Syunyaev}}]{1983ASPRv...2..189P}
{Pozdnyakov}, L.~A., {Sobol}, I.~M., \& {Syunyaev}, R.~A. 1983, \apspr, 2, 189

\bibitem[{{Rybicki} \& {Lightman}(1979)}]{1979rpa..book.....R}
{Rybicki}, G.~B., \& {Lightman}, A.~P. 1979, {Radiative processes in
  astrophysics}

\bibitem[{{Schnittman} {et~al.}(2013{\natexlab{a}}){Schnittman}, {Krolik}, \&
  {Noble}}]{schnittmanetal2013}
{Schnittman}, J.~D., {Krolik}, J.~H., \& {Noble}, S.~C. 2013{\natexlab{a}},
  \apj, 769, 156, \dodoi{10.1088/0004-637X/769/2/156}

\bibitem[{{Schnittman} {et~al.}(2013{\natexlab{b}}){Schnittman}, {Krolik}, \&
  {Noble}}]{schnittman2013}
---. 2013{\natexlab{b}}, \apj, 769, 156, \dodoi{10.1088/0004-637X/769/2/156}

\bibitem[{{Shakura} \& {Sunyaev}(1973)}]{shakura_sunyaev1973}
{Shakura}, N.~I., \& {Sunyaev}, R.~A. 1973, \aap, 500, 33

\bibitem[{{S{\k{a}}dowski} \& {Narayan}(2016)}]{sadowski_narayan2016}
{S{\k{a}}dowski}, A., \& {Narayan}, R. 2016, \mnras, 456, 3929,
  \dodoi{10.1093/mnras/stv2941}

\bibitem[{{S{\k{a}}dowski} {et~al.}(2015){S{\k{a}}dowski}, {Narayan},
  {Tchekhovskoy}, {Abarca}, {Zhu}, \& {McKinney}}]{sadowskietal2015}
{S{\k{a}}dowski}, A., {Narayan}, R., {Tchekhovskoy}, A., {et~al.} 2015, \mnras,
  447, 49, \dodoi{10.1093/mnras/stu2387}

\bibitem[{{S{\k{a}}dowski} {et~al.}(2013){S{\k{a}}dowski}, Narayan,
  Tchekhovskoy, \& Zhu}]{sadowskietal2013}
{S{\k{a}}dowski}, A., Narayan, R., Tchekhovskoy, A., \& Zhu, Y. 2013, Monthly
  Notices of the Royal Astronomical Society, 429, 3533,
  \dodoi{10.1093/mnras/sts632}

\bibitem[{{Skinner} {et~al.}(1982){Skinner}, {Bedford}, {Elsner}, {Leahy},
  {Weisskopf}, \& {Grindlay}}]{skinneretal1982}
{Skinner}, G.~K., {Bedford}, D.~K., {Elsner}, R.~F., {et~al.} 1982, \nat, 297,
  568, \dodoi{10.1038/297568a0}

\bibitem[{Skinner \& Ostriker(2013)}]{skinneretal2013}
Skinner, M.~A., \& Ostriker, E.~C. 2013, The Astrophysical Journal Supplement
  Series, 206, 21, \dodoi{10.1088/0067-0049/206/2/21}

\bibitem[{{Socrates} \& {Davis}(2006)}]{socrates_davis2006}
{Socrates}, A., \& {Davis}, S.~W. 2006, \apj, 651, 1049, \dodoi{10.1086/507119}

\bibitem[{{Spitzer}(1956)}]{spitzer1956}
{Spitzer}, L. 1956, {Physics of Fully Ionized Gases}

\bibitem[{{Steiner} {et~al.}(2009){Steiner}, {Narayan}, {McClintock}, \&
  {Ebisawa}}]{steineretal2009}
{Steiner}, J.~F., {Narayan}, R., {McClintock}, J.~E., \& {Ebisawa}, K. 2009,
  \pasp, 121, 1279, \dodoi{10.1086/648535}

\bibitem[{{Stepney}(1983)}]{stepney1983}
{Stepney}, S. 1983, \mnras, 202, 467, \dodoi{10.1093/mnras/202.2.467}

\bibitem[{{Stone} {et~al.}(2008){Stone}, {Gardiner}, {Teuben}, {Hawley}, \&
  {Simon}}]{stoneetal2008}
{Stone}, J.~M., {Gardiner}, T.~A., {Teuben}, P., {Hawley}, J.~F., \& {Simon},
  J.~B. 2008, \apjs, 178, 137, \dodoi{10.1086/588755}

\bibitem[{{Stone} {et~al.}(1992){Stone}, {Mihalas}, \&
  {Norman}}]{stoneetal1992}
{Stone}, J.~M., {Mihalas}, D., \& {Norman}, M.~L. 1992, \apjs, 80, 819,
  \dodoi{10.1086/191682}

\bibitem[{{Stone} {et~al.}(2020){Stone}, {Tomida}, {White}, \&
  {Felker}}]{stoneetal2020}
{Stone}, J.~M., {Tomida}, K., {White}, C.~J., \& {Felker}, K.~G. 2020, \apjs,
  249, 4, \dodoi{10.3847/1538-4365/ab929b}

\bibitem[{{Straub} {et~al.}(2011){Straub}, {Bursa}, {S{\k{a}}dowski},
  {Steiner}, {Abramowicz}, {Klu{\'z}niak}, {McClintock}, {Narayan}, \&
  {Remillard}}]{straubetal2011}
{Straub}, O., {Bursa}, M., {S{\k{a}}dowski}, A., {et~al.} 2011, \aap, 533, A67,
  \dodoi{10.1051/0004-6361/201117385}

\bibitem[{Sądowski \& Narayan(2015)}]{sadowski_narayan2015}
Sądowski, A., \& Narayan, R. 2015, Monthly Notices of the Royal Astronomical
  Society, 453, 3213, \dodoi{10.1093/mnras/stv1802}

\bibitem[{Sądowski {et~al.}(2014)Sądowski, Narayan, McKinney, \&
  Tchekhovskoy}]{sadowskietal2014}
Sądowski, A., Narayan, R., McKinney, J.~C., \& Tchekhovskoy, A. 2014, Monthly
  Notices of the Royal Astronomical Society, 439, 503,
  \dodoi{10.1093/mnras/stt2479}

\bibitem[{Tully {et~al.}(2013)Tully, Courtois, Dolphin, Fisher, Héraudeau,
  Jacobs, Karachentsev, Makarov, Makarova, Mitronova, Rizzi, Shaya, Sorce, \&
  Wu}]{tullyetal2013}
Tully, R.~B., Courtois, H.~M., Dolphin, A.~E., {et~al.} 2013, The Astronomical
  Journal, 146, 86, \dodoi{10.1088/0004-6256/146/4/86}

\bibitem[{Turner \& Stone(2001)}]{turnerstone2001}
Turner, N.~J., \& Stone, J.~M. 2001, The Astrophysical Journal Supplement
  Series, 135, 95, \dodoi{10.1086/321779}

\bibitem[{{Verner} {et~al.}(1996){Verner}, {Ferland}, {Korista}, \&
  {Yakovlev}}]{verneretal1996}
{Verner}, D.~A., {Ferland}, G.~J., {Korista}, K.~T., \& {Yakovlev}, D.~G. 1996,
  \apj, 465, 487, \dodoi{10.1086/177435}

\bibitem[{{Walton} {et~al.}(2016){Walton}, {Middleton}, {Pinto}, {Fabian},
  {Bachetti}, {Barret}, {Brightman}, {Fuerst}, {Harrison}, {Miller}, \&
  {Stern}}]{Walton2016}
{Walton}, D.~J., {Middleton}, M.~J., {Pinto}, C., {et~al.} 2016, \apjl, 826,
  L26, \dodoi{10.3847/2041-8205/826/2/L26}

\bibitem[{{Walton} {et~al.}(2018){Walton}, {F{\"u}rst}, {Heida}, {Harrison},
  {Barret}, {Stern}, {Bachetti}, {Brightman}, {Fabian}, \&
  {Middleton}}]{Walton2018_sample}
{Walton}, D.~J., {F{\"u}rst}, F., {Heida}, M., {et~al.} 2018, \apj, 856, 128,
  \dodoi{10.3847/1538-4357/aab610}

\bibitem[{{Walton} {et~al.}(2020){Walton}, {Pinto}, {Nowak}, {Bachetti},
  {Sathyaprakash}, {Kara}, {Roberts}, {Soria}, {Brightman}, {Canizares},
  {Earnshaw}, {F{\"u}rst}, {Heida}, {Middleton}, {Stern}, {Tao}, {Webb},
  {Alston}, {Barret}, {Fabian}, {Harrison}, \& {Kosec}}]{waltonetal2020}
{Walton}, D.~J., {Pinto}, C., {Nowak}, M., {et~al.} 2020, \mnras, 494, 6012,
  \dodoi{10.1093/mnras/staa1129}

\bibitem[{{Webb} {et~al.}(2017){Webb}, {Gu{\'e}rou}, {Ciambur}, {Detoeuf},
  {Coriat}, {Godet}, {Barret}, {Combes}, {Contini}, {Graham}, {Maccarone},
  {Mrkalj}, {Servillat}, {Schroetter}, \& {Wiersema}}]{webbetal2017}
{Webb}, N.~A., {Gu{\'e}rou}, A., {Ciambur}, B., {et~al.} 2017, \aap, 602, A103,
  \dodoi{10.1051/0004-6361/201630042}

\bibitem[{{White} {et~al.}(2023){White}, {Mullen}, {Jiang}, {Davis}, {Stone},
  {Morozova}, \& {Zhang}}]{white2023}
{White}, C.~J., {Mullen}, P.~D., {Jiang}, Y.-F., {et~al.} 2023, arXiv e-prints,
  arXiv:2302.04283, \dodoi{10.48550/arXiv.2302.04283}

\bibitem[{{White} {et~al.}(2016){White}, {Stone}, \& {Gammie}}]{whiteetal2016}
{White}, C.~J., {Stone}, J.~M., \& {Gammie}, C.~F. 2016, \apjs, 225, 22,
  \dodoi{10.3847/0067-0049/225/2/22}

\bibitem[{Wibking \& Krumholz(2022)}]{wibkingkrumholz2022}
Wibking, B.~D., \& Krumholz, M.~R. 2022, Monthly Notices of the Royal
  Astronomical Society, 512, 1430, \dodoi{10.1093/mnras/stac439}

\bibitem[{{Wilms} {et~al.}(2000){Wilms}, {Allen}, \& {McCray}}]{wilmsetal2000}
{Wilms}, J., {Allen}, A., \& {McCray}, R. 2000, \apj, 542, 914,
  \dodoi{10.1086/317016}

\bibitem[{{Zhu} {et~al.}(2012){Zhu}, {Davis}, {Narayan}, {Kulkarni}, {Penna},
  \& {McClintock}}]{zhuetal2012}
{Zhu}, Y., {Davis}, S.~W., {Narayan}, R., {et~al.} 2012, \mnras, 424, 2504,
  \dodoi{10.1111/j.1365-2966.2012.21181.x}

\end{thebibliography}

\end{document}